\documentclass[useAMS,amssymb,usegraphicx,usenatbib,twocolumn,revtex4-1]{aastex6} 
\usepackage{aas_macros}
 
\usepackage[english]{babel}     
\usepackage{enumitem,amssymb,amsmath,graphicx,braket,hyperref,natbib,bm,multirow}                      

\def\aj{AJ}

\def\apj{ApJ}
\def\apjl{ApJ}
\def\apjs{ApJS}
\def\aap{A\&A}
\def\mnras{MNRAS}
\def\nat{Nature}
\def\na{NewA}
\def\pasp{PASP} 
\def\pasa{PASA} 

\def\aaps{A\&AS} 

\makeatother

\shorttitle{Building the peanut}
\shortauthors{Saha et al.}

\begin{document}

\title{Building the peanut: simulations and observations of peanut-shaped
  structures and ansae in face-on disk galaxies}

\author{Kanak Saha}
\affil{Inter-University Centre for Astronomy and Astrophysics, Pune-411007, India.}
\email{kanak@iucaa.in} 
\author{Alister W.\ Graham and Isabel Rodr\'{\i}guez-Herranz}
\affil{Centre for Astrophysics and Supercomputing, Swinburne University of Technology, Victoria 3122, Australia.}

\begin{abstract}
(X/peanut)-shaped features observed in a significant fraction of disk galaxies 
are thought to have formed from vertically buckled bars. Despite being three dimensional 
structures, they are preferentially detected in near edge-on projection.
Only a few galaxies are found to have displayed such structures when their 
disks are relatively face-on - suggesting that either they are generally 
weak in face-on projection or many may be hidden by the light of their galaxy's 
face-on disk. 

Here we report on three (collisionless) simulated galaxies displaying peanut-shaped structures 
when their disks are seen both face-on and edge-on - resembling a three-dimensional peanut or 
dumbbell. Furthermore, these structures are accompanied by ansae and an outer ring at the end
of the bar --- as seen in real galaxies such as IC~5240.

The same set of quantitative parameters used to measure peanut
structures in real galaxies have been determined for the simulated
galaxies, and a broad agreement is found. In addition, the peanut
length grows in tandem with the bar, and is a maximum at half the
length of the bar. Beyond the cutoff of these peanut structures,
towards the end of the bar, we discover a new positive/negative
feature in the $B_6$ radial profile associated with the isophotes of
the ansae/ring.

Our simulated, self-gravitating, three-dimensional peanut structures
display cylindrical rotation even in the near-face-on disk
projection. In addition, we report on a kinematic pinch in the
velocity map along the bar minor-axis, matching that seen in the
surface density map.

\end{abstract}


\section{Introduction}
\label{sec:intro}

Disk galaxies host ``bulges'' with diverse morphologies and kinematics. While classical 
bulges appear rounder, and are somewhat dominated by random stellar motion (although see 
\cite{Sahaetal2012}), ``pseudobulges'' may have boxy/peanut-shapes or even display an X-shaped 
morphology, and they are typically supported by ordered rotational motion \citep{Combes2011,
Athanassoula2016}. \cite{Butaetal2010} correctly refers to these latter structures as differing 
bar morphology rather than bulge morphology. More than $50\%$ of edge-on disk galaxies in the 
local universe are thought to host these (boxy/peanut)-shaped features \citep{BureauFreeman1999,
Luttickeetal2000a,Luttickeetal2000b,YoshinoYamauchi2015,ErwinDebattista2017}, as does the Milky Way 
and M31 \citep{Dweketal1995, AthanassoulaBeaton2006, Ciamburetal2017}.  It should be noted that 
the presence of a ``pseudobulge'', which is related to the presence of a bar, does not come at 
the expense of a classical bulge, which are typically smaller in size. When the disks of galaxies 
hosting these (X/peanut/bowtie)-shaped features are viewed face-on, these features are thought to 
appear oval in shape, and have been referred to as ``barlenses'' \citep{Laurikainenetal2011,
Laurikainenetal2014,Athanassoulaetal2015,LaurikainenSalo2017}.

The pioneering $N$-body work by \cite{CombesSanders1981}, and subsequent studies by 
\cite{Combesetal1990, Rahaetal1991}, demonstrated that such peanut structures can be  
formed via the vertical buckling instability of a stellar bar --- such that the inner 
part of the bar thickens in the vertical direction and when the disk is viewed edge-on, 
with the bar perpendicular to the line-of-sight, a distinct boxy/peanut, or at times 
X-shaped, morphology appears. Such a morphological structure has been shown to have 
drawn its support from an orbital backbone associated with vertical inner Lindblad 
resonance through which star particles are excited in the vertical direction, i.e.\ 
out of the disk plane \citep{Pfenniger1985, PfennigerFriedli1991,Patsisetal2002,MV2006}. 
Several $N$-body simulations have confirmed this scenario for peanut-shaped structure 
formation \citep{PfennigerNorman1990, Rahaetal1991, Athanamisi2002, MV2004,Athanassoula2005,
Debattistaetal2006,Sahaetal2013}. The association of these peanut structures with bars has 
additionally been established kinematically \citep{KuijkenMerrifield1995,BureauFreeman1999,
ChungBureau2004,Debattistaetal2005,Williamsetal2011}. Even relatively face-on galaxies, 
e.g.\ NGC~98 $i=40\deg$), have kinematic signatures of such peanut structures 
\citep{Mendez-Abreuetal2008}, and recently \cite{ErwinDebattista2016} presented further evidence, 
based on IFU kinematics, of a buckling instability leading to the formation of the boxy/peanut
structures previously seen in the somewhat face-on galaxies NGC~3227 
($i=48\deg$, \cite{Laurikainenetal2011}) and NGC~4569 ($i=62\deg$, \cite{Laurikainenetal2004, 
Jogeeetal2005}). Collisionless $N$-body simulations by \cite{Athanamisi2002,ONeilDubinski2003} 
have shown such peanut structures in face-on projection.

Although not well known, the appearance of such boxy/peanut-shaped structures is {\it not} 
limited to the observed edge-on galaxies. That is, galaxies seen with rather face-on disks 
can also present peanut structures, rather than just oval-shaped barlenses.  IC~5240
\citep{Buta1995}, see also \cite{Laurikainenetal2011}, and to a lesser extent
IC~4290 \citep[their figure~12 reveals what they identify as boxiness in the inner part 
of the bar]{ButaCrocker1991} are two examples, as are NGC~4123 and NGC~4314 \citep{Blocketal2001}, 
and the {\it Third Reference Catalogue of Bright Galaxies} \citep{deVaucouleurs1991} even
contains a (poorly known) discussion of such galaxies. Further examples can be found 
in \cite{Quillenetal1997,Laurikainenetal2011,ErwinDebattista2013}. A point of difference 
to note here is that we do not consider these to just be projections of a 2D bowtie/peanut 
structure existing in the $z$-direction of the galaxy (e.g.\ \cite{Debattistaetal2005}), 
but rather are structures in the $x-y$ disk plane. The identification of these peanut 
signatures in galaxies with rather face-on disks raises a number of questions regarding 
their formation as pointed out by \cite{Laurikainenetal2011} who found nine local galaxies 
with a thickened peanut-shaped inner bar and disk inclinations less than 65 degrees.

\begin{figure*}
\includegraphics[width=1.0\textwidth]{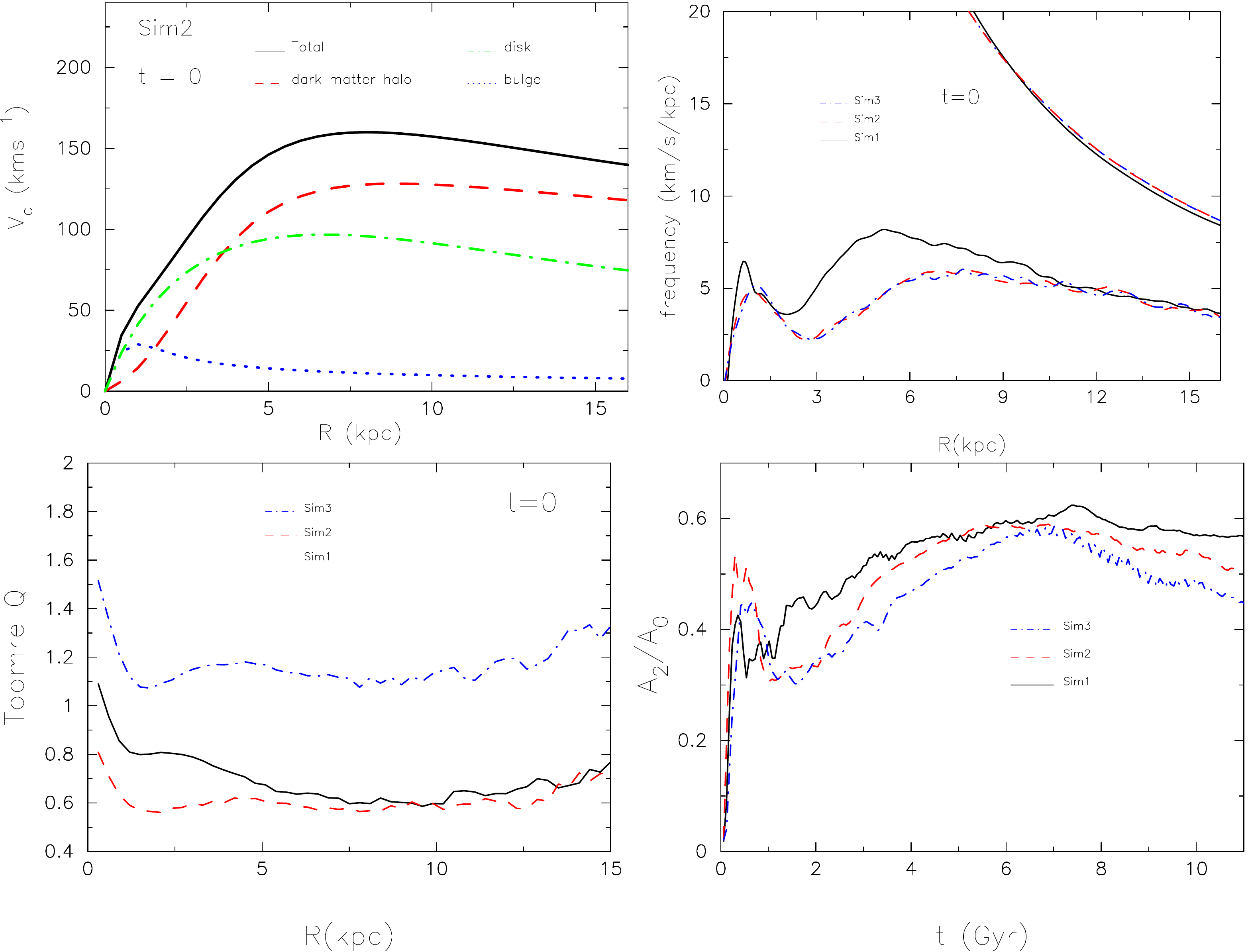}
\caption{Top left: Initial ($t=0$) circular velocity curves for Sim2. 
Top right: Initial radial variation of 
$\Omega -\kappa/2$ (curves with double humps) and $\Omega$. 
Bottom left: Initial radial variation
of Toomre $Q$ parameter. 
Bottom right: Time evolution of the bar strength as measured by the maximum
 of the m=2 Fourier component in the particle distribution.}
\label{fig:VcQA2}
\end{figure*}

In this paper, we report on the formation of strong peanut structures that are visible 
in simulations of face-on disk galaxies created using collisionless $N$-body simulations 
and compare these structures with the prototype ``face-on peanut'' galaxy IC~5240, plus 
several ``edge-on peanuts''. Building on past studies of bar strength, 
\citep[e.g.][her figure~4]{Combes2016}, and following \cite{CG2016}, we provide a
quantitative structural analysis of the peanut structures using Fourier
harmonics to describe the deviations from pure ellipses of a) the isodensity
contours, in the case of the simulations, and b) the surface brightness maps,
in the case of the real galaxies.  A bar introduces a strong m=2 Fourier mode, while a 
bar plus a boxy/peanut structure can be represented by an m=6 perturbation \citep{CG2016}. 
The strength of this perturbation, and its spatial dimensions, can be measured in the same
manner for both simulations and real galaxy images. Having such a set of ``peanut
parameters'' not only enables us to better compare the models with real
galaxies, but has the promise of enabling real galaxies to be matched against
the evolutionary tracks of simulated galaxies in various parameter scaling
diagrams, thereby enabling one to age-date the peanut structures observed in
real galaxies. 

The rest of the paper is organized as follows: in section~\ref{sec:modelsetup}, we present 
the setup of the disk galaxy models, explain the ensuing $N$-body simulation, and reveal 
the emergence of the peanut structures which are visible in the face-on orientation of 
the disks. In section~\ref{sec:method}, we more fully describe the metrics used to quantify 
peanut structures. From the Fourier analysis, we present five quantitative measures
of the peanut, such as its length and strength \citep{CG2016}.  We then apply
this in section~\ref{sec:application} to our set of three simulated galaxies,
and, for the first time, to a peanut observed in a real galaxy whose disk
appears somewhat face-on.  In section~\ref{sec:compare} we compare the
structural parameters of both the simulated and real peanuts, and 
we present the evolution of a simulated galaxy in these parameter
scaling diagrams. Section~\ref{sec:bkin} presents the kinematics of the
simulated peanut structure, as seen when the host disk is both face-on and
edge-on. Finally, a discussion and our primary conclusions are presented in
section~\ref{sec:discussion}.

\section{The Simulation}
\label{sec:modelsetup}

\subsection{Initial Setup} 

We present three equilibrium galaxy simulations (Sim1, Sim2 and Sim3), each
consisting of an initially axisymmetric disk, a dark matter halo, and a small
classical bulge. The initial stellar disk surface density follows an
exponential profile along the radial direction, and an approximately
$sech^2(z)$ profile along the vertical direction. The initial dark matter
distribution is modeled as a cored halo giving rise to a nearly flat rotation
curve in the outer parts \citep{Evans1993}, and the initial bulge is modeled
with a cored King profile \citep{King1966}. Each component in the model is
constructed using a distribution function (DF, which is a function of
integrals of motion, satisfying the collisionless Boltzmann equation) and are
live, allowing them to interact with each other. Further details on the
model construction can be found in \cite{KD1995, Sahaetal2012}. The
structural properties of the stellar disk are identical in all three
models. The mass models differ in terms of their spheroidal components -- the
bulge and halo component in Sim1 is slightly more massive than those in Sim2
and Sim3, see Table~\ref{mass}.

\begin{table}
\begin{center}
\caption{Galaxy mass model} 
\label{mass}
\begin{tabular}{l c c c c }
\hline
Model & $M_{bulge}$ & $M_{disk}$  & $M_{halo}$ & $Q$\\
    & ($\times 10^{10}M_{\odot}$) & ($\times 10^{10}M_{\odot}$)& ($\times 10^{10}M_{\odot}$) & ($R=R_d$) \\
\hline
Sim1  & 0.032  & 1.67 & 6.8 & 0.80\\ 
Sim2  & 0.022  & 1.67 & 6.5 & 0.60\\
Sim3  & 0.022  & 1.67 & 6.5 &1.15\\
\hline
\end{tabular}
\end{center}
$Q$ is the Toomre parameter. 
$R_d$is the initial disk scalelength, equal to 3~kpc.
\end{table}

We have scaled the models such that the initial disk mass $M_{disk} = 1.67
\times 10^{10}~M_{\odot}$, and the initial disk scalelength $R_d=3.0$ kpc.  
The time between two consecutive snapshots is $60$~Myr. Thus, snapshot 0 marks $t=0$ 
and snapshot 200 marks $t=200=12$~Gyr. 
The circular velocity curve for Sim2 is shown in Fig.\ref{fig:VcQA2}.  The
radial variation of $\Omega - \kappa/2$ for the three models are shown in the
top right panel of Fig.\ref{fig:VcQA2}; here, $\Omega$ is the circular
frequency of the stars in the disk, and $\kappa$ is their radial epicyclic frequency.  
As combined above, $\Omega - \kappa/2$ refers to the free precession frequency
of an $m=2$ mode in the absence of self-gravity and stellar pressure \citep{BT1987}.
The difference seen here between both Sim2 and Sim3, when compared with Sim1, can be attributed to the
different bulge-to-disk mass ratios.  The bottom left panel of
Fig.\ref{fig:VcQA2} depicts the initial Toomre Q variation with radius.  Both
Sim1 and Sim2 have $Q < 1$ at nearly all radii, whereas in Sim3 $Q > 1$ at all
radii.

We have used a total of $3.7$ million particles, with $0.5 \times 10^6$ in the
bulge, $1.2 \times 10^6$ in the disk, and $2.0\times 10^6$ in the dark
halo. The masses of the bulge, disk and halo particles are $451.3 M_{\odot}$, $1.4\times 10^{4} M_{\odot}$
and $3.1 \times 10^{4} M_{\odot}$ respectively. The softening lengths for the disk, bulge and halo particles are chosen
following the suggestion of \cite{McMillan2007} for unequal mass particles. The simulations were
performed using the Gadget-1 code \citep{Springeletal2001} with a tolerance parameter $\theta_{tol} =0.7$, 
integration time step (advancing particle coordinates) $\sim 1.8$ Myr. The simulations were evolved for a total 
time period of $\sim 13.2$ Gyr. The total energy is conserved within $2\%$ for the entire run. For Sim1, energy is conserved within $1\%$. 
The angular momentum conservation, however, is not good enough for such a long run; the total angular 
momentum is
conserved within $5.0\%$. We have checked that full system (bulge+disk+halo) is in virial equilibrium
with $T/|W| \sim 0.5$, ($T, W$ are total kinetic and potential energy) maintained with deviation $< 0.1 \%$.

\subsection{Evolution}

\begin{figure*}
{\includegraphics[width=1.0\textwidth ]{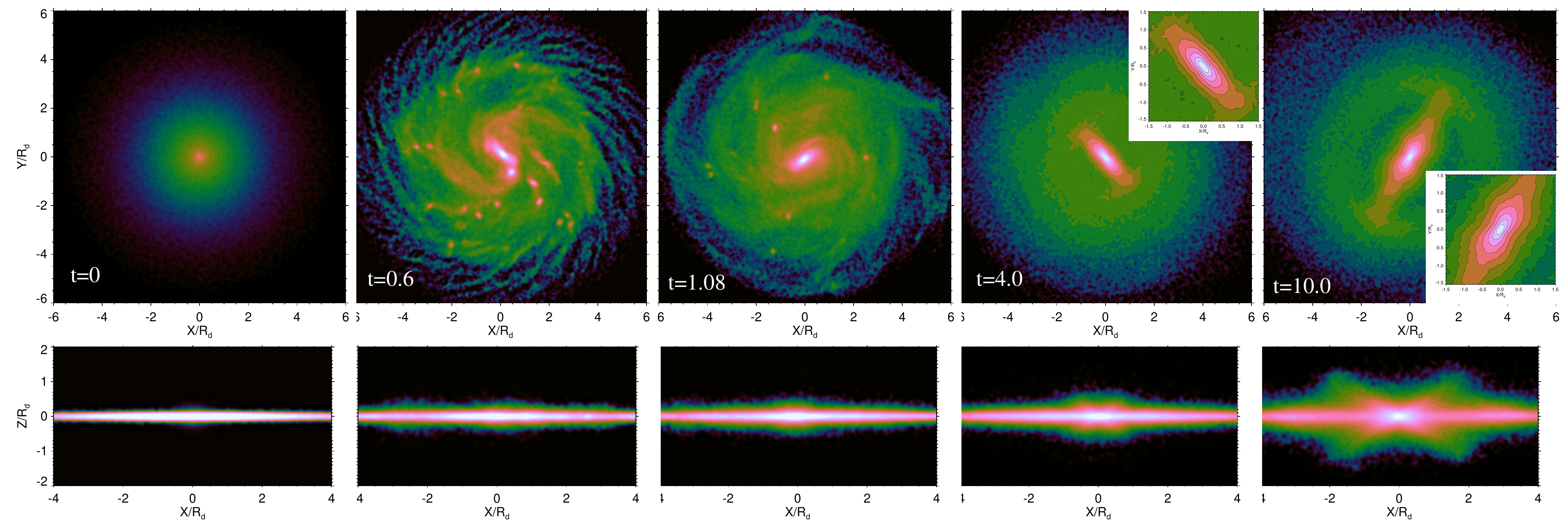}}
\caption{Top panel: Face-on surface density maps at different epochs for
  Sim2. The inset images display the isodensity contours for the inner region of the model. 
Bottom panel: Edge-on maps at the same epochs. The unit of time is in Gyr.}
\label{fig:img-seq}
\end{figure*}

\setcounter{figure}{2}
\begin{figure*}
\includegraphics[width=0.8\textwidth]{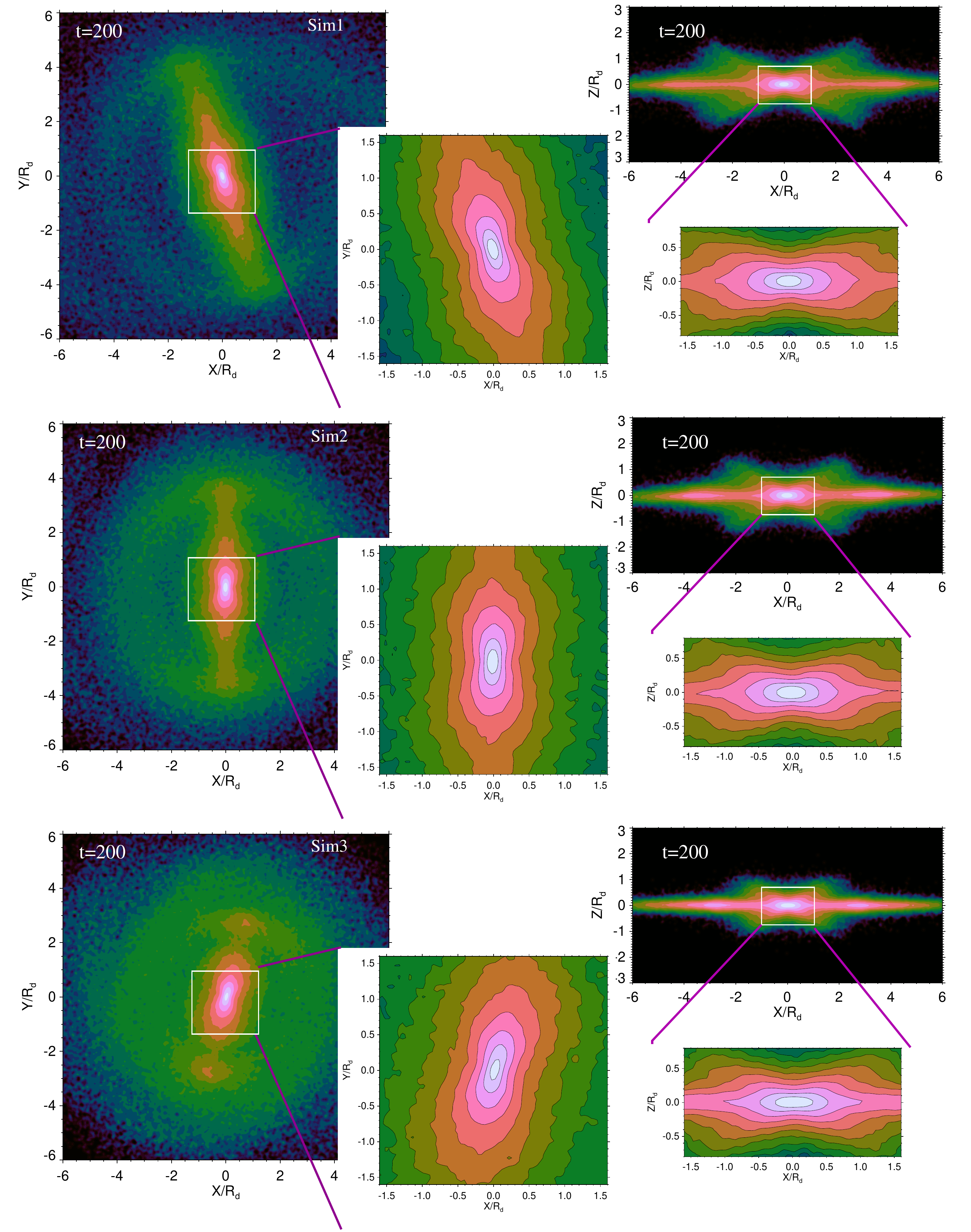}
\caption{Top panels: Sim1 at time step $t=200$ (12.0 Gyr).  
Both a face-on ($i=0$~deg) and an edge-on ($i=90$~deg) view of the disk
reveals the peanut morphology. 
For the edge-on projection, the major-axis of the bar has been rotated and
aligned with the X-axis of the image. 
The inner 1.6 $R_d$ (4.8 kpc) region shown by the zoomed-in image displays a clear peanut
structure in both the Y-X and Z-X plane, while the fainter and more extended 
(X/bowtie)-shaped structure appears in the Z-X plane.
Middle panels: same as above but for Sim2. 
Bottom panels: same as above but for Sim3. Note: The white boxes in each panel
are not to scale, but are purely illustrative.} 
\label{fig:t200}
\end{figure*}

\begin{figure*}
\begin{center}
\includegraphics[width=0.7\textwidth]{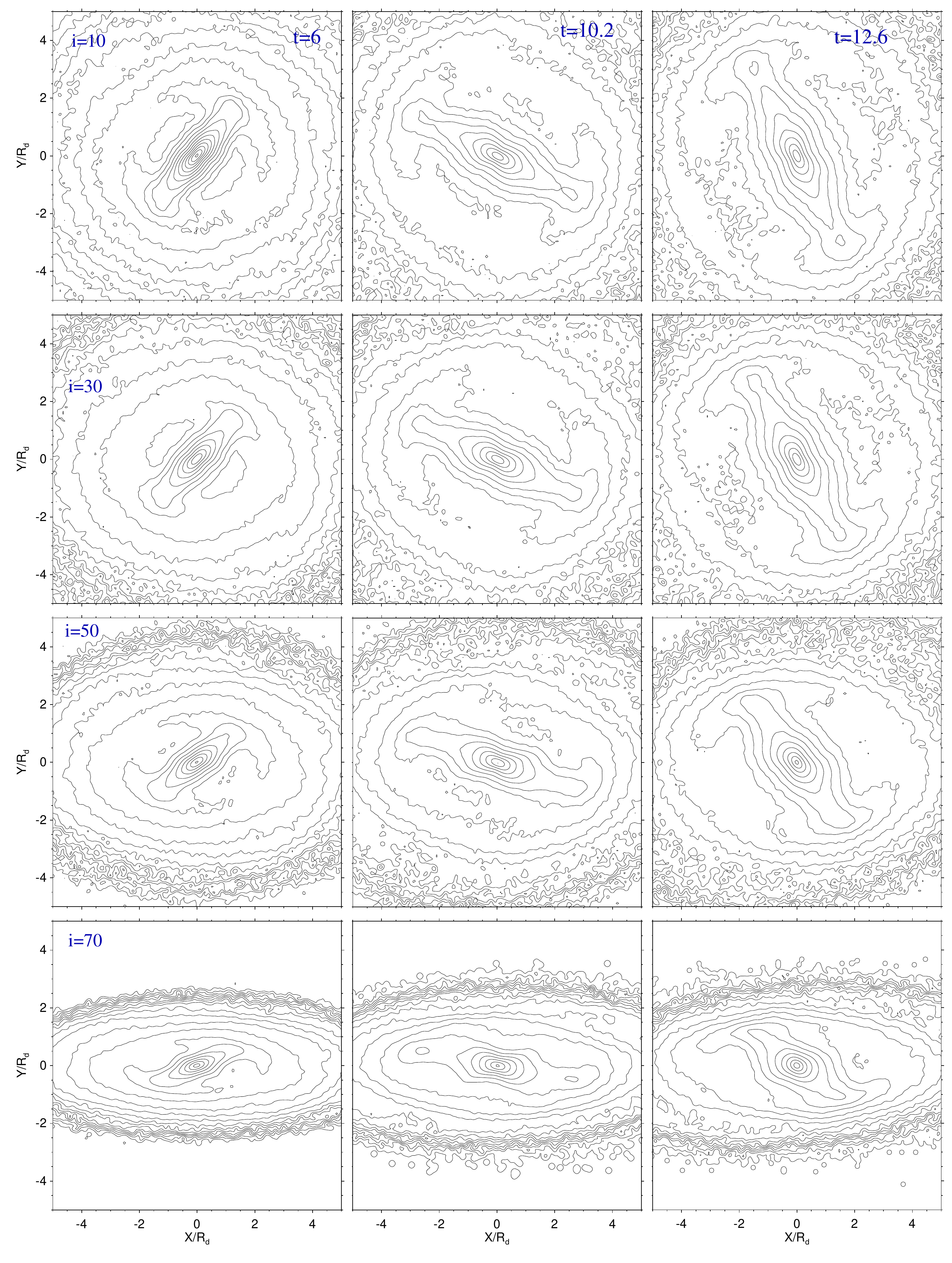}
\caption{Surface density contours for Sim2, projected with different 
inclination angles, and at three different epochs (6, 10.2 and 12.6 Gyr).} 
\label{fig_contour}
\end{center}
\end{figure*}

The initial axisymmetric stellar disks in Sim1 and Sim2, being cold (with $Q <
1$ in most of the disk), are subjected to the strongest disk instability
due to an axisymmetric mode ($m=0$ Fourier mode) \cite{Toomre1969}. As a result, 
the stellar disks undergo fragmentation in the early phase --- leading to the 
formation of stellar clumps which migrate to the center and enrich the pre-existing 
classical bulge, see Fig.~\ref{fig:img-seq}, similar to what might be happening in
high redshift gas rich galaxies having giant star-forming clumps \citep{Elmegreenetal2008}.
The migration and coalescence of the stellar clumps in our simulations
occur on a rather faster time scale; by about $2$~Gyr, most clumps have migrated
to the center. The effect of this on the bulge growth can be appreciated
visually by inspecting the edge-on images in Fig.~\ref{fig:img-seq} at $t=0$
and $t=1.08$~Gyr. These stellar disks also form short-lived, multi-arm,
spiral structures which dissolve, resulting in heating of the stellar disk
\citep{Sahaetal2010,SellwoodCarlberg2014}. While the stellar disk in Sim3,
being cool (rather than cold), having $Q > 1$, does not go through any fragmentation. It evolves
by forming multi-arm spiral structures which also end up heating the stars in
the disk.

All three stellar disks form a strong bar ($m=2$ Fourier component) in the
early phase, reaching a peak amplitude of about $A_{2}/A_{0} \sim 0.5$
within $ t \sim 0.6$ Gyr. The bar amplitude then decays for roughly 1~Gyr,
before rising again (Fig.~\ref{fig:VcQA2}). Note that the initial decay of
the bar amplitude is not due to the well-known buckling instability
\citep{MV2006} but to the fact that those stellar clumps migrate to the center
due to dynamical friction from all directions and merge with the existing bar,
leading to a weakening of the bar strength. After this brief period of minimum,
the bar starts growing again, and over the next several billion years the
stellar disk builds an unusually long bar, especially in Sim1 and Sim2 (with
initial $Q < 1$), see Fig.~\ref{fig:img-seq}.

\subsection{Emergence of the peanut and ansae} 

In each simulation, the bar undergoes a buckling instability and forms a
boxy/peanut structure which becomes apparent at around $4$~Gyr. The strength
of this structure increases as time progresses, and it eventually evolves into
an X-shaped morphology when seen from the edge-on projection.  This
boxy/peanut feature is not only visible in the edge-on projection of the disk
galaxy (lower panel of Fig.~\ref{fig:img-seq}), but also in the face-on
projection (upper panel of Fig.~\ref{fig:img-seq}). The isophotes of the
central region, within about $1.6~R_d$ (4.8 kpc), are highlighted with the inset images in
the top panel of Fig.~\ref{fig:img-seq}.

Expanding on the previous observation, Fig.~\ref{fig:t200} clearly reveals the
peanut structure at $t=12$~Gyr, both in the face-on and edge-on projection,
from all three of our simulations --- implying that the ``peanut'' is actually a
three-dimensional (3D) structure common to all the models presented here, 
confirming previous collisionless simulations showing 
face-on peanuts \citep{Athanamisi2002,ONeilDubinski2003}. 

In addition to this boxy/peanut structure, we observe the development of two other
structures. In the edge-on view of the galaxy, the peanut extends to larger
radii by morphing into a bowtie or X-shaped structure which continues until a
certain radial extent, as seen in the X-Z plane. This radial cutoff is
associated with the onset of the partial spiral ring or ansae seen at the end
of the bar in the face-on projection of the galaxy.  Particles/stars at the
end of the bar do not exit the disk plane to form the X/Peanut structure, but
are instead confined to the disk plane and help to build the ansae/ring.

It remains to be fully investigated what causes such 3D peanut structures. 
Our simulations do not appear to support a trend of face-on peanut
formation with Toomre Q. Sim3 has a Toomre Q that is a factor of 2 higher than
in Sim1 and Sim2, yet Sim3 still has a similar degree of face-on peanut.  We
ran an additional simulation with a slightly hotter stellar disk $Q \sim 1.5$
(cf. 1.15 in Sim 3) 
and it still formed a 3D peanut.  To emphasis that these 3D peanuts are not
due to any projection effects, in Fig.~\ref{fig_contour} we show the density
contours from Sim2 taken at four different inclination angles ($i=10^{\circ},
30^{\circ}, 50^{\circ}$ and $70^{\circ}$), and at 3 different epochs during
the evolution.  At $t=6$~Gyr, there is no sign of a face-on peanut at any of
the inclination angles. At $t=10.2$ and $12.6$~Gyr, the morphology of the
contours reveal the presence of a boxy/peanut structure.  At lower inclination
angles ($0\deg$ being face-on), there is a large-scale bar with a clear pinch along the bar minor
axis, and an associated partial outer ring --- an unambiguous signature of a
face-on peanut.

In the following sections we quantify these face-on peanut structures, as well
as that seen in the galaxy IC~5240.

\section{Quantifying (face-on) peanuts}
\label{sec:method}

To the best of our knowledge, peanut structures seen in face-on disks have
never been quantified before. 
We follow the procedure introduced by \cite{CG2016} to quantitatively describe
the properties of peanut-shaped structures seen in edge-on disk galaxies.  The
distinction here is that we are not measuring a vertical off-plane
($z$-direction) feature, but rather an in-plane ($x-y$) feature. The procedure is
capable of detecting the existence of peanut features even when they are not
very prominent nor the dominant component in the inner region of a galaxy 
(where they can coexist with the bar, classical bulge and disk).  As briefly
noted in the Introduction, the technique uses 
Fourier harmonics to describe the deviations of a galaxy's isophote from pure
ellipses.  For a circle of radius $R$, the (azimuthal angle)-dependent
radius of the quasi-circle is given by 
\begin{equation}
  R'(\theta) = R + \sum_{n} \left[ A_{n} \textrm{sin}(n\theta) + B_{n}
    \textrm{cos}(n\theta) \right] ,
  \label{equ:FHarmpsi}
\end{equation}
where $\theta$ is the azimuthal angle, and $n$ is the harmonic order. 

\begin{figure}
\includegraphics[width=0.5\textwidth]{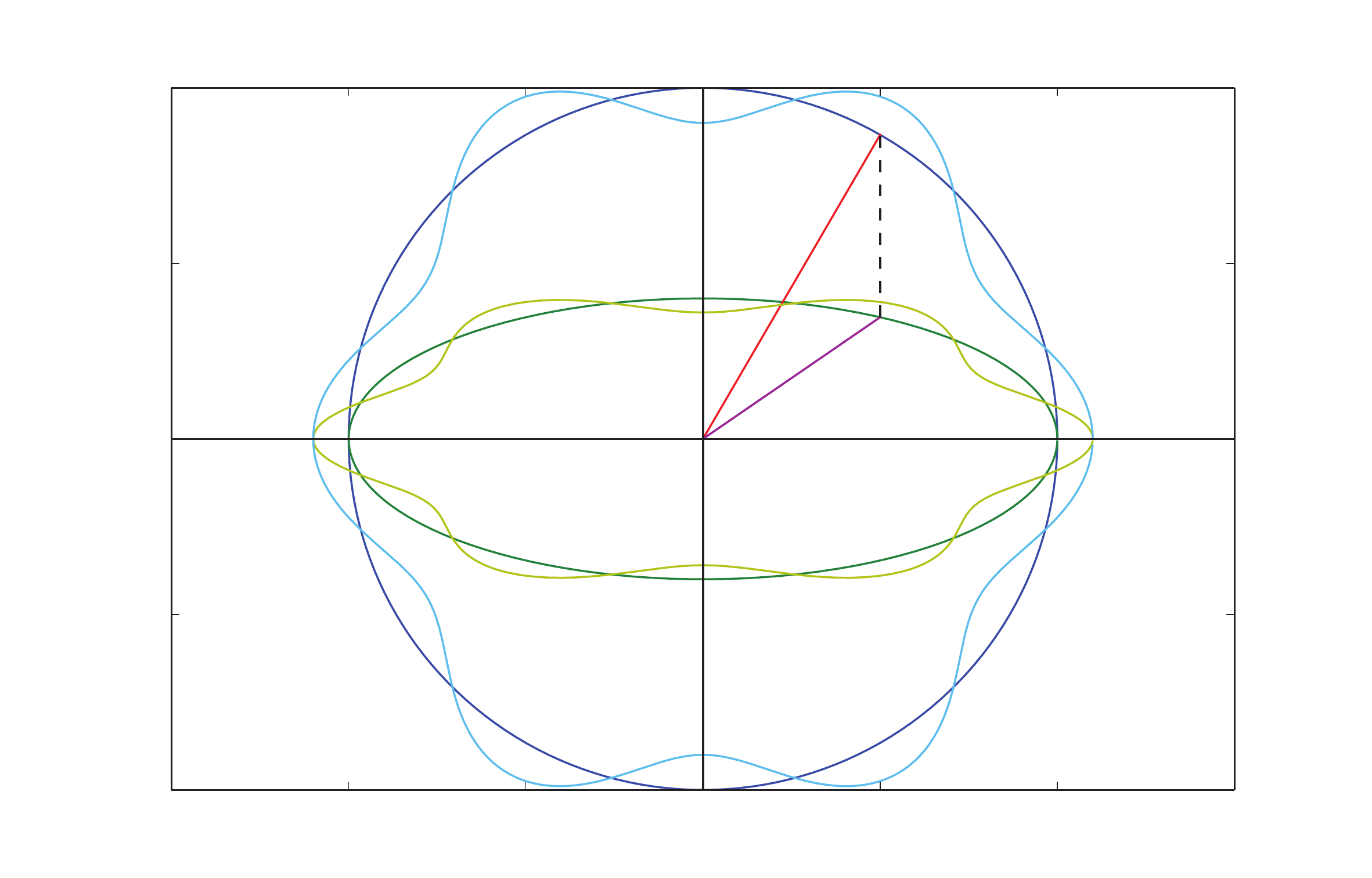}
\caption{The dark green ellipse with ellipticity $e \equiv 1 - b/a = 0.6$ is distorted
  using the sixth order cosine term $B_6 \cos(6\psi)$ of the Fourier harmonic
  series to create the light green curve. In this example, $B_6 = 0.1$ has been
  used.  Such a curve is representative of the isophotes of a peanut plus a bar
  within a face-on disk. The angle $\psi$ is the eccentric anomaly of the
  ellipse.  In this example, we see how an azimuthal angle $\theta =
  60^{\circ}$ on the circle, marked by the red line's departure from the
  positive X-axis, maps to the reduced angle
  shown by the purple line.  The light blue curve shows the $B_6
  \cos(6\theta)$ deviations.}
\label{fig:Fig_FH}
\end{figure}

While a positive $m=2$ cosine harmonic can represent the isophotal
perturbation of a bar in a face-on disk, and a negative $m=4$ cosine mode can
reproduce (boxy/peanut)-shaped isophotes, it is a positive $m=6$ cosine mode
which captures the combination of a (boxy/peanut)-shaped feature plus a longer
bar seen within a face-on disk.  This sixth order cosine term introduces six
positive and six negative deviations, from a pure ellipse, as one traverses
each ellipse in azimuthal angle (see Figure~\ref{fig:Fig_FH}). For a galaxy
whose disk is viewed with an edge-on orientation, the six positive deviations
match up with the four ``prongs'' of the peanut structure plus the two
``prongs'' of the edge-on disk.  
With either an edge-on disk or a face-on bar
contributing positively at an azimuthal angle of 0 and 180 degrees, the
positive deviations for the peanut feature would be located at azimuthal
angles of $\alpha=$ 60$^{\circ}$ above and below the semi-major axis for an
isophote with ellipticity $e=0$, i.e.\ for a circle. If $e \neq 0$, these
angles are distorted by the quotient of the minor-to-major axis $b/a$ (see
Figure~\ref{fig:Fig_FH}).

As one steps out in radius, each quasi-elliptical isophote has its own
amplitude of the $B_6$ term used for quantifying the strength of these
deviations.  From this radial $B_{6}$ profile, one can measure the five
parameters defined in section 2 of \cite{CG2016}.  Aware that these ``peanut parameters'' are
relatively new in the literature, and therefore likely to be unfamiliar to
some readers, we summarize them here. \\ 
\textbf{(a)} \textbf{$\Pi_{max}$}: the maximum amplitude of the $B_{6}$
   coefficient from all radii. \\
\textbf{(b)} \textbf{$R_{\Pi,max}$}: the radial length of the peanut
structure at which $\Pi_{max}$ occurs, as
   projected onto the major-axis. \\
\textbf{(c)} \textbf{$xy_{\Pi,max}$}: the extent of the peanut perpendicular
  to the bar, but in the disk plane\footnote{\cite{CG2016} defined a parameter
  called $z_{\Pi,max}$ because they studied edge-on disk galaxies for which they
  computed the vertical extent of the peanut structures. In our analysis, we
  explore an equivalent metric in the $xy$ disk plane of the galaxy, as we are
  examining face-on disk systems.\label{foot1}}. \\
\textbf{(d)} \textbf{$S_{\Pi}$}: the 'strength' of the peanut, given by the integral
   between radii at which $B_{6} = \Pi_{max}/2$. \\
\textbf{(e)} \textbf{$W_{\Pi}$}: the width of the peanut's $B_6$ profile, 
 given by the distance between the previous two radii.\\ 
Readers may like to refer to Figures~2 and 3 from \cite{CG2016} for a visual
description of these parameters.

Our images were modeled using the new tasks {\sc Isofit} and {\sc Cmodel} developed for
the IRAF software package. {\sc Isofit} was developed by \cite{C15Isofit} to
significantly enhance the modeling of galaxies whose isophotes (or isodensity
contours in the case of simulations) depart from elliptical shapes.
Taking into account all relevant Fourier terms, {\sc Isofit} varies the intensity of
the light in the model to achieve the best match to the 2D light distribution.
One can then build a 1-D surface brightness profile along the major-axis,
$\mu(R_{\rm maj})$, that includes all of these deviations from a pure
ellipse.  {\sc Isofit} provides the radial profile of the harmonic terms necessary to
do this.  

We performed a decomposition of the extracted, major-axis, surface brightness
profiles using {\sc Profiler} \citep{Ciambur2016}, enabling us to quantify the
different constituents of which the systems are composed. For our
investigation of the surface brightness profile, the primary component for the
analysis is the disk and its inferred scale-length.  This is employed to
normalize the previous ``peanut parameters'' in order to perform a fair
comparison among different systems, regardless of their distance or size.  This
decomposition additionally allows us to determine how the length of the
peanut, determined by the $B_6$ profile, correlates with the length of the
bar, determined from the surface brightness profile.

\section{Application}
\label{sec:application}

In this section, we will analyze the face-on projections of the three different simulated
galaxies, and the real galaxy IC~5240 ($i_{\rm disk}=49^{\circ}$, where
0$^{\circ}$ refers to the face-on view). Unfortunately, we found that our method
struggled to model the edge-on projections of the simulations because the
X-shaped features do not lie at the appropriate azimuthal angles, and the
spikes of the X-shape were too narrow, to be properly captured using Fourier harmonics. 
This prevented us from obtaining a clean run of the $B_6$ profile. However,
as our investigation pertains to face-on peanuts, our focus was on the face-on
projections and we were able to model these well. 

For the galaxy IC~5240, we will analyze a 3.6 $\mu m$ Spitzer image \citep{Shethetal2010}. 
While modeling the surface brightness profile, the fitted model components were convolved 
with the Point Spread Function (PSF), which was estimated with the IRAF task imexamine by 
fitting a Moffat profile to 10 stars in the field.

The simulated and real galaxy in our sample do not exhibit a single
continuous ellipticity profile, $\epsilon(R)$, but instead possess an abrupt
transition/jump from high to low values of ellipticity once the bar ends and
the ring appears. Thus, we ran {\sc Isofit} two consecutive times. First, for the
inner region, we let all the parameters (ellipticity, position angle, and
center coordinates) free. Second, from the end of the bar to the outskirts of the
galaxy, we fixed the position angle to that of the bar. The best solution for
all the galaxies is obtained by using Fourier terms 2, 3, 4, 6, 8, and 10 in
{\sc Isofit}.  A discussion of the $m=8$ mode is beyond the scope of this study, but
shall be presented in a forthcoming paper, while a physical meaning for the
(very weak) $m=10$ mode remains elusive.

\begin{figure}
\includegraphics[width=\columnwidth]{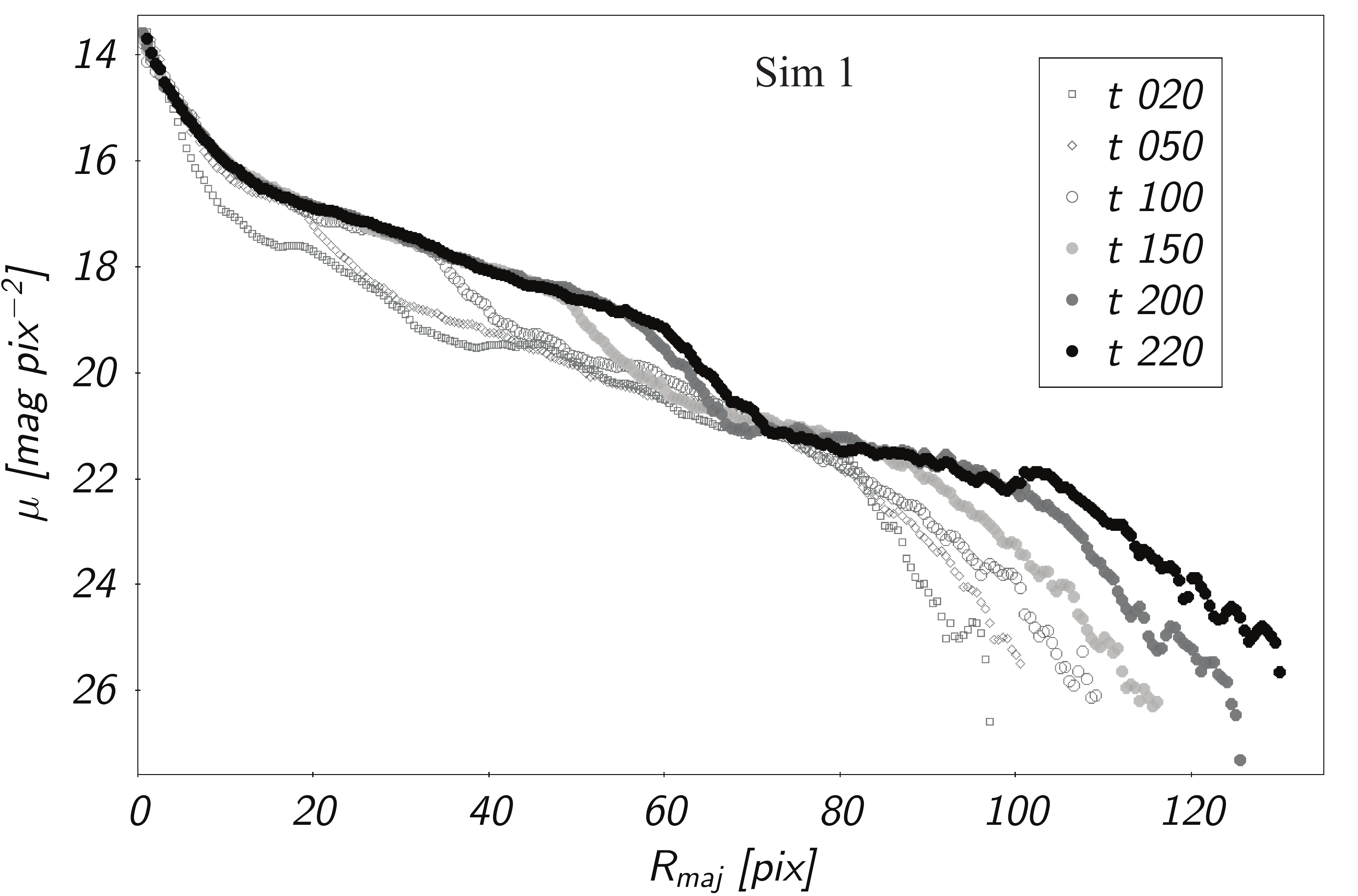}
\caption{Major-axis surface brightness profiles for Sim1, shown at six time steps: 
$t=$20, 50, 100, 150, 200, 220. A single time step $\Delta t = 1$ equals 60 Myr.  
The growth of the bar can readily be seen after $t=20$ (1.2 Gyr). 
For all simulations: 1 pix=1 arcsec=0.234 kpc.}
\label{fig_SBPall}
\end{figure}

\begin{figure}
\includegraphics[width=\columnwidth]{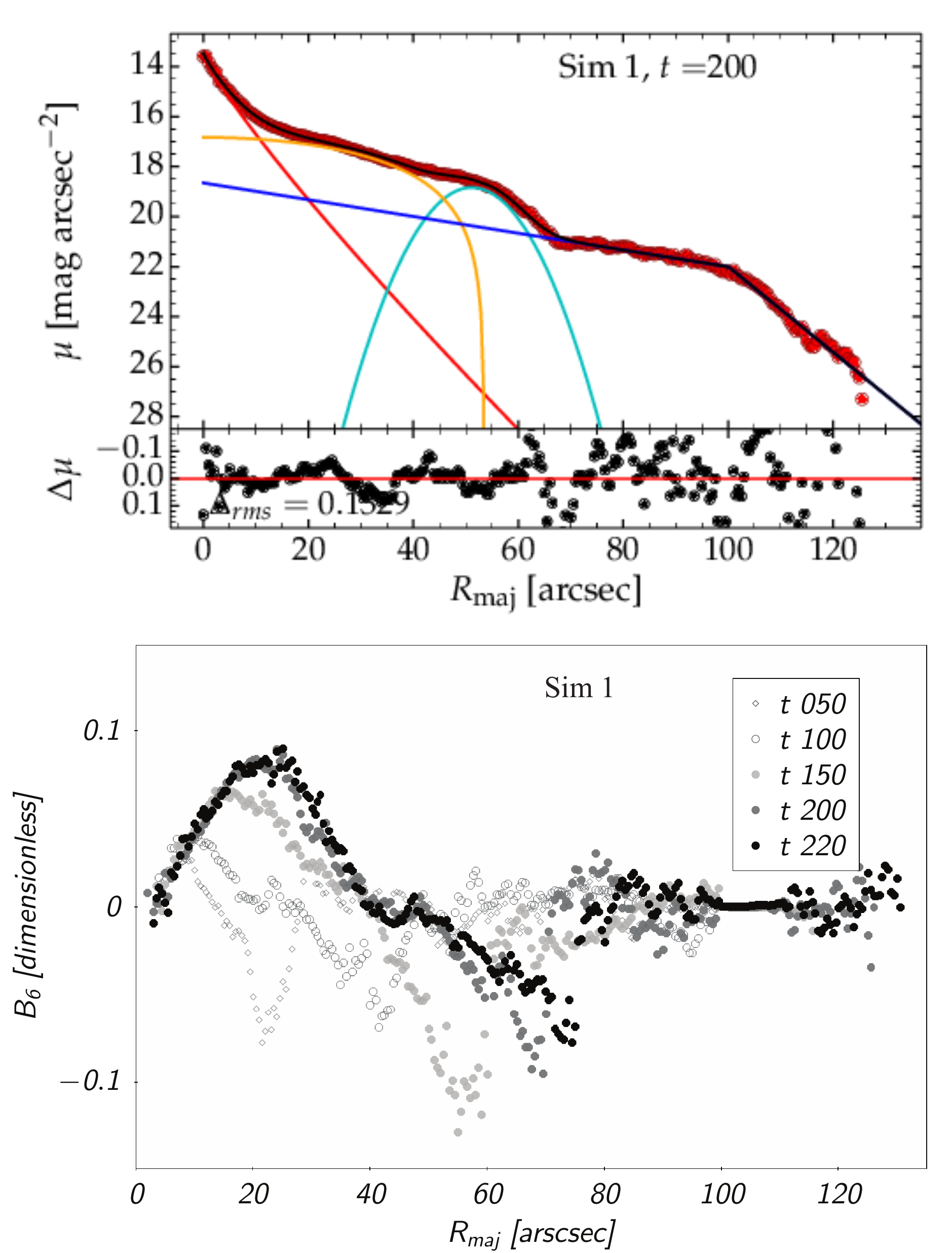}
\caption{Top: Decomposition of the semi-major axis, surface brightness profile
  of Sim1, when $t=200$ (12 Gyr): classical S\`ersic bulge with $n=1.2$ (red), broken exponential
  disk (dark blue), Ferrers bar (orange), and Gaussian ansae/ring (cyan).  The
  radial scale is such that one arcsecond equals one pixel=0.234 kpc. The
  inner disk scale-length $h \approx 32\arcsec \approx 7.5$~kpc. 
  The residual profile of the model subtracted from the data is also shown.
  Bottom: $B_{6}$ radial profiles for simulation 1 at time steps $t=$50, 100,
  150, 200, 220 ($=13.2$~Gyr) show the growth of both the peanut (+ve $B_6$) and
  the ansae at the end of the bar (small +ve bump then large -ve dip in the
  $B_6$ profile).}
\label{fig_B6all}
\end{figure}

\subsection{The face-on peanut in Sim1}

The top panel of Fig.~\ref{fig:t200} presents our first simulation, at a time
step $t=200$ (12 Gyr) --- the peanut structure is evident in both the edge-on and
the face-on projections of the disk. Fig.~\ref{fig_SBPall} and
\ref{fig_B6all} reveal the evolution of the bar and associated peanut
structure, since its formation at an early age ($t=20$ for the bar, and $t=50$
for the peanut) until the bar reaches a size of roughly half the stellar disk
at 12 Gyr.
Figures~\ref{fig_SBPall} and \ref{fig_B6all} display the major-axis surface
brightness profile and the $B_6$ profile, as measured from the face-on
orientation of the disk. In Figure~\ref{fig_B6all}, the distance from the
center of the galaxy, along the major-axis, has been expressed in arcseconds, such
that one arcsec corresponds to one pixel in the simulation. 
As the peanut grows with time, the amplitude and width of the associated bump
in the $B_6$ profile can be seen to increase. The negative dips in the $B_6$
profiles are due to the ansae at the end of the bar.  
Following section~\ref{sec:method}, we have computed the five parameters that
characterize the ``peanutness'' of the galaxy using the positive peak in the
$B_6$ profile. The results are gathered in Table~\ref{tab:param1}.

\begin{table}
\begin{center}
\caption{Peanut Parameters: Sim1} 
\label{tab:param1}
\begin{tabular}{l c c c c c}
\hline 
time                & 50    & 100   & 150   & 200   & 220 \\
\hline
$\Pi_{max}$         & 0.03  & 0.04  & 0.07  & 0.09  & 0.09\\
\hline 
$R_{\Pi,max}$ [kpc] & 0.91  & 1.29  & 1.94  & 3.07  & 3.19\\
$xy_{\Pi,max}$ [kpc]    & 0.54  & 0.70  & 1.05  & 1.55  & 1.62\\
$S_{\Pi}$ [kpc]     & 2.13  & 3.80  & 12.90  &17.0  & 18.65\\
$W_{\Pi}$ [kpc]     & 0.73  & 1.19  & 2.48  & 2.65  & 2.65\\
\hline
\end{tabular}
\end{center}

Peanut parameters for Sim1, as seen with a face-on disk, at time steps $t=$50,
100, 150, 200, 220 ($=13.2$~Gyr), expressed in kpc, except for $\Pi_{max}$
which is a dimensionless quantity. The inner exponential disk scale-length was
found to be fairly constant at $\approx$7.5 kpc prior to disk
bending/truncation at large radii.
\end{table}

Having quantified the evolution of the peanut in Sim1, we are able to make a
number of observations. Over time, the length of the bar grows, as does the
radial length where the peanut's presence is a maximum, as denoted by
$R_{\Pi,max}$.  The ratio of the peanut length to the bar length was observed
to remain roughly constant at $0.5$ as the bar and peanut co-evolved. The
length of the peanut and the bar --- relative to the near constant exponential
disk scalelength --- are thus an indication of the age of these structures in
Sim1.

As the bar length and peanut length grow in tandem, the amplitude of the
maximum $B_6$ Fourier cosine term used to identify the peanut, denoted by
$\Pi_{\max}$, was also observed to increase, therefore making it easier to
identify the peanut (relative to the background disk) as time
increased. Consequently, the bar strength, measured by $S_{\Pi}$, also grows
stronger with time.

In contrast to the above evolution, the ratio of the peanut's height relative
to its length, i.e.\ $xy_{\Pi,max}/R_{\Pi,max}$, was found to be fairly constant
with time, decreasing only slightly from 0.59 at $t=50$ to 0.51 at $t=220$.
The width of the peanut's hump in the $B_6$ profile relative to the length of
the peanut, i.e.\ $W_{\Pi}/R_{\Pi,max}$ also appears steady at around 0.8 to
0.9 (with the exception of a departure to a ratio of $\approx$1.3 at $t=150$). Therefore,
the $W_{\Pi}/xy_{\Pi,max}$ ratio was also fairly constant at around 0.6--0.75
(with the same exception of the $t=150$ snapshot).

In the top panel of Fig.~\ref{fig_B6all}, the optimal decomposition of the
surface brightness profile of Sim1 at $t=$200 (12 Gyr) is shown.  It was
derived using the {\sc Profiler} software \citep{Ciambur2016}. The
decomposition consists of an $n=1.2$ S\`ersic function to describe the
classical bulge,
a double exponential function for the disk with an inner scale-length of
$h=7.5$~kpc, a Ferrers function for the bar and a Gaussian function for the
ansae/ring at the end of the bar. The length of the bar at $\approx 45\arcsec$
($\approx 10.5$~kpc) is twice the distance where the peanut feature in the
$B_6$ profile is a maximum at $\approx 23\arcsec$ ($\approx 5.4$~kpc). This
same ratio has been observed in real disk galaxies viewed edge-on
\citep{Luttickeetal2000a,LaurikainenSalo2017,ErwinDebattista2017}.

It should be noted that, along the major-axis, the peanut-shaped  structure or
``pseudobulge'' does not contribute much signal relative to the remaining
galaxy light, and is effectively subsumed back into the bar component.  That
is, there is no additional component required for the peanut in the
decomposition.  This holds true for all the simulations studied here, and for
IC~5240, and also for all of the edge-on galaxies with peanut structures
presented in \cite{CG2016}.  We suspect that it is highly probable that many
past studies of galaxies claiming to have identified and modeled a
``pseudobulge'' have in fact identified and quantified the classical bulge
(which can have S\'ersic indices $n < 2$).  For a discussion and understanding
of this topic, see \cite{GrahamConf2014, GrahamConf2015}. 

\subsection{The face-on peanuts in Sim2 and Sim3}

\begin{figure}
\includegraphics[width=\columnwidth]{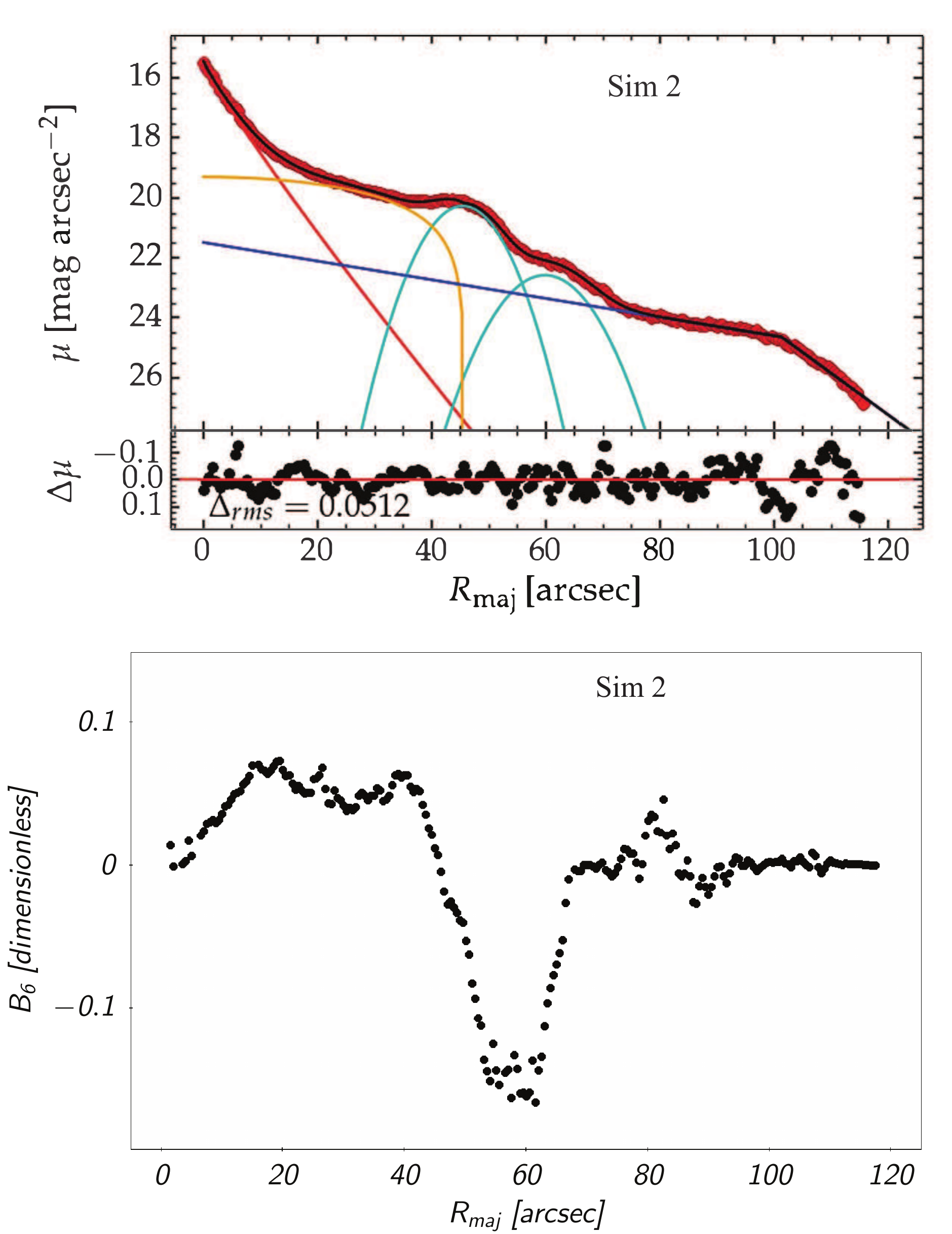}
\caption{Top: Decomposition of the major-axis surface brightness profile, and
  residual profile (data-model), of Sim2 at $t=200$ ($12$~Gyr), into a
  classical S\`ersic
  bulge with $n=0.8$ (red line), double exponential disk (dark blue), Ferrers bar
  (orange), Gaussian ansae at $\approx$45$\arcsec$, and Gaussian ring at
  $\approx$60$\arcsec$ (both in cyan).  The radial scale is such that one
  arcsecond equals one pixel (=0.234 kpc). The inner disk scale-length 
  $h=34.9$ arcsec.  Bottom: The associated $B_{6}$ radial profile.  The first
  positive bump is due to the peanut; the second bump then subsequent dip are
  due to the ansae.}
\label{fig-Decom-B6-Sim2}
\end{figure}

\begin{figure}
\includegraphics[width=\columnwidth]{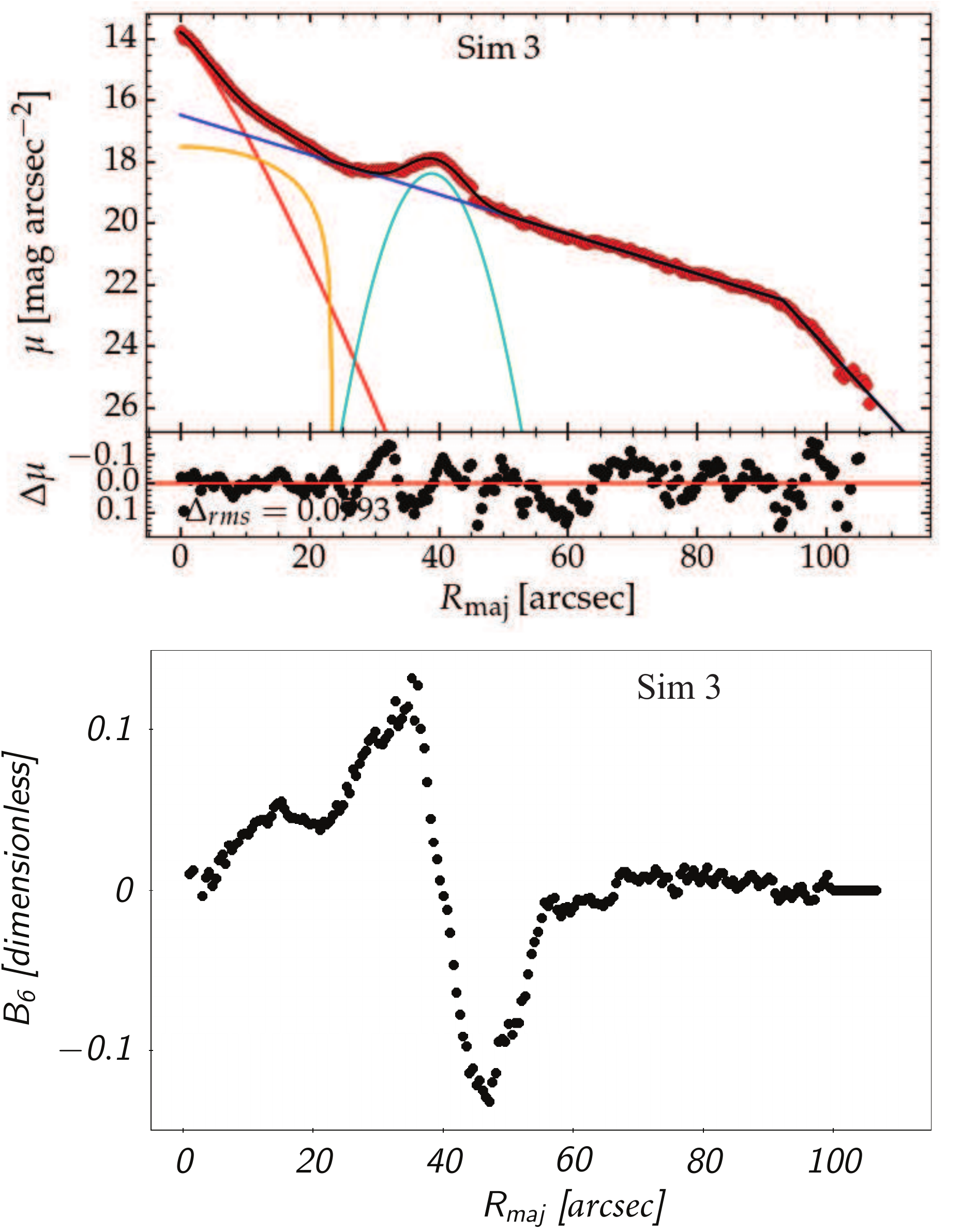}
\caption{Top: Decomposition of the major-axis surface brightness profile, and
  residual profile (data-model), of Sim3 at $t=200$ (=$12$~Gyr), into a
  classical S\`ersic bulge with $n=1.1$ 
  (red line), double exponential disk (dark blue), Ferrers bar (orange), and Gaussian ansae.
 (cyan). The radial scale is such that one
  arcsecond equals one pixel (=0.234 kpc). Bottom: The associated $B_{6}$ radial profile.}
\label{fig-Decom-B6-Sim3}
\end{figure}

The top panels in Figures~\ref{fig-Decom-B6-Sim2} and
\ref{fig-Decom-B6-Sim3} present the major-axis surface brightness profiles
at $t=200$ (12~Gyr) for the models Sim2 and Sim3, respectively. These
have been fit with galaxy components. Both decompositions consist of a
classical S\`ersic bulge ($n=0.8$ and $n=1.1$, respectively), a Ferrers bar, 
a double (or broken) exponential disk, and 
additional Gaussians that account for the ansae and rings at the end of the
bar, features common to barred galaxies \citep{MVKB07}.

The bottom panels of Figures~\ref{fig-Decom-B6-Sim2} and
\ref{fig-Decom-B6-Sim3} present the $B_6$ radial profile of each
simulation. The first positive bump corresponds to the peanut feature, while
the second positive bump together with the negative dip are due to the ansae.
As in the previous simulation, the peanut is seen to peak at roughly half the
length of the bar. The radius where the maximum of the $B_6$ peak occurs is
close to $20\arcsec$=$4.68$~kpc for Sim2 and about $15\arcsec$=$3.5$~kpc for Sim3, whilst their bars
end at $\sim 9.2$~kpc and $\sim 6.9$~kpc, respectively.

The maximum amplitude of the $B_6$ profile from these face-on peanut structures at this time of the evolution
are $\Pi_{max}=0.09, 0.07$, and $0.06$ for Sim1, Sim2 and Sim3 respectively.
The peanut parameters derived from the analysis of these two model galaxies
are presented in Table~\ref{tab:param2}.
What is also apparent from looking at the face-on disk projection in
Fig.~\ref{fig:t200}, and from the $B_6$ profiles in 
Figs.~\ref{fig_B6all}--\ref{fig-Decom-B6-Sim3}, is that a more prominent
ansae/pseudo-ring at the end of the bar is associated with a stronger peak in
the $B_6$ profile, which is then followed at larger radii by a negative dip.
This will be further investigated elsewhere.

\subsection{The face-on peanut in galaxy IC~5240}

The same methodology as above has been used here to characterize the peanut in the
real galaxy IC~5240.  The distance to IC~5240 is 26.9$\pm$1.9 Mpc according to
NED\footnote{NASA/IPAC Extragalactic Database: \url{ned.ipac.caltech.edu}},
based on Plank cosmology \citep{Planck2016}, and applying the Virgo, Great
Attractor, and Shapley motion corrections \citet{Mouldetal2000} available
through NED. For IC~5240, one arcsecond equals 0.13 kpc.

\begin{figure}
\includegraphics[width=0.8\columnwidth]{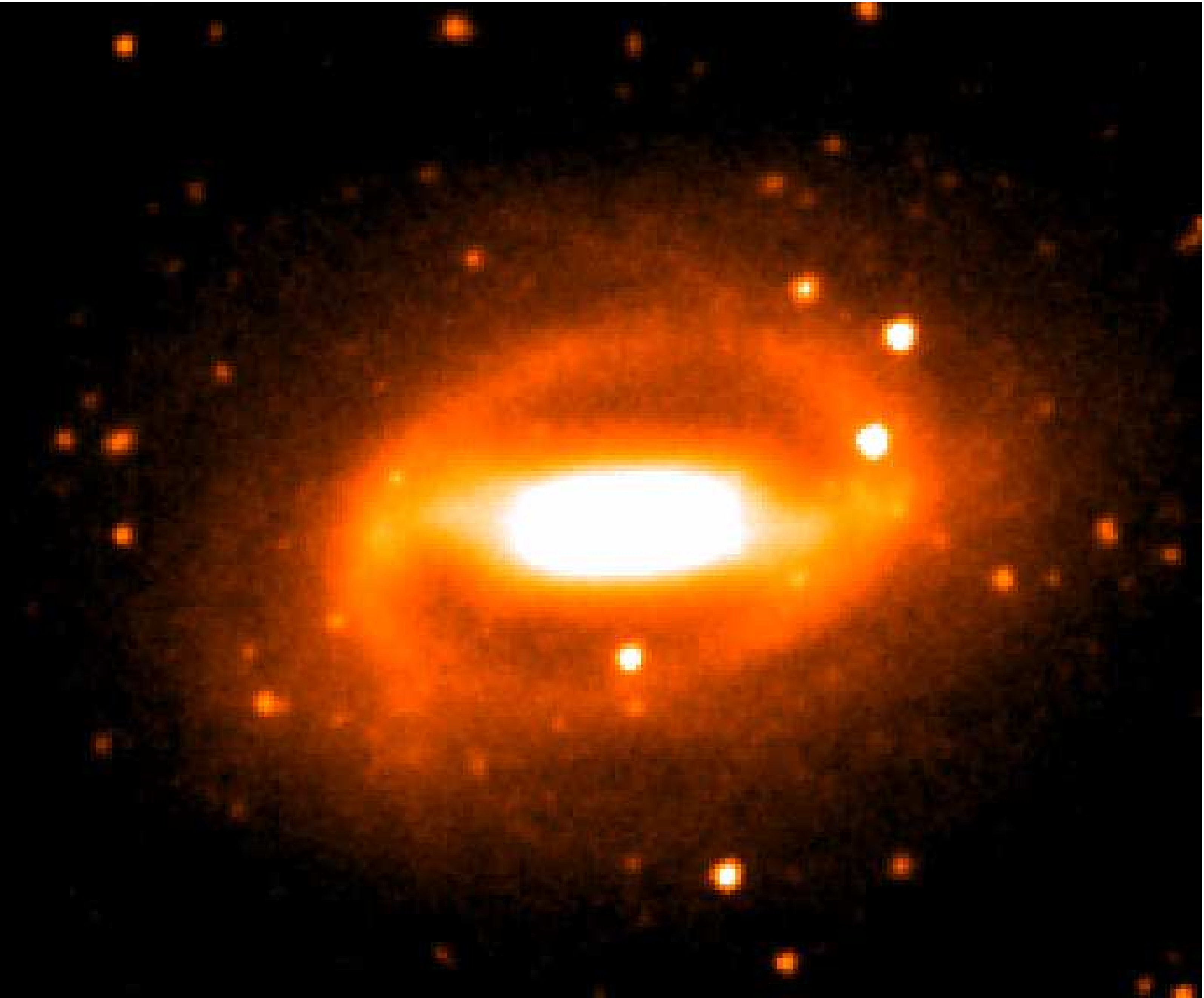}
\caption{Image of the real galaxy IC~5240 observed in 3.6 $\mu m$ band by the
  Spitzer Space Telescope.}
\label{fig:IC5240}
\end{figure}

\begin{figure}
\includegraphics[angle=0,  width=\columnwidth]{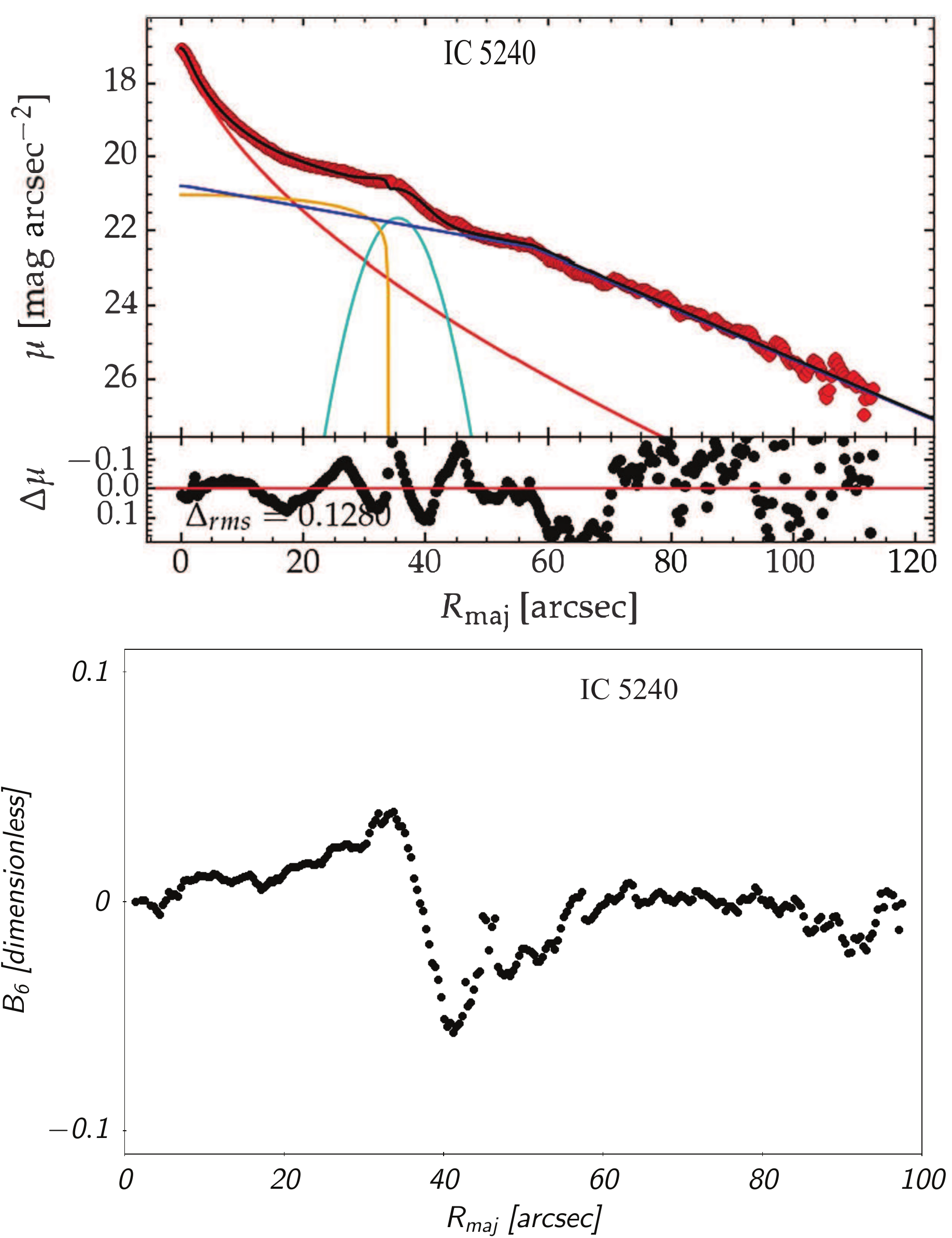}
\caption{Top: Decomposition of the major-axis surface brightness profile, and
  residual profile (data-model), of galaxy IC~5240 into a classical S\`ersic
  bulge with $n=1.2$ (red line), double exponential disk (dark blue), Ferrers
  bar (orange), and Gaussian ansae/ring (cyan). Bottom: The associated $B_{6}$
  radial profile.  $1\arcsec = 0.13$~kpc.}
\label{Decom-B6-IC5240}
\end{figure}

\begin{figure}
\includegraphics[height=7.0 cm]{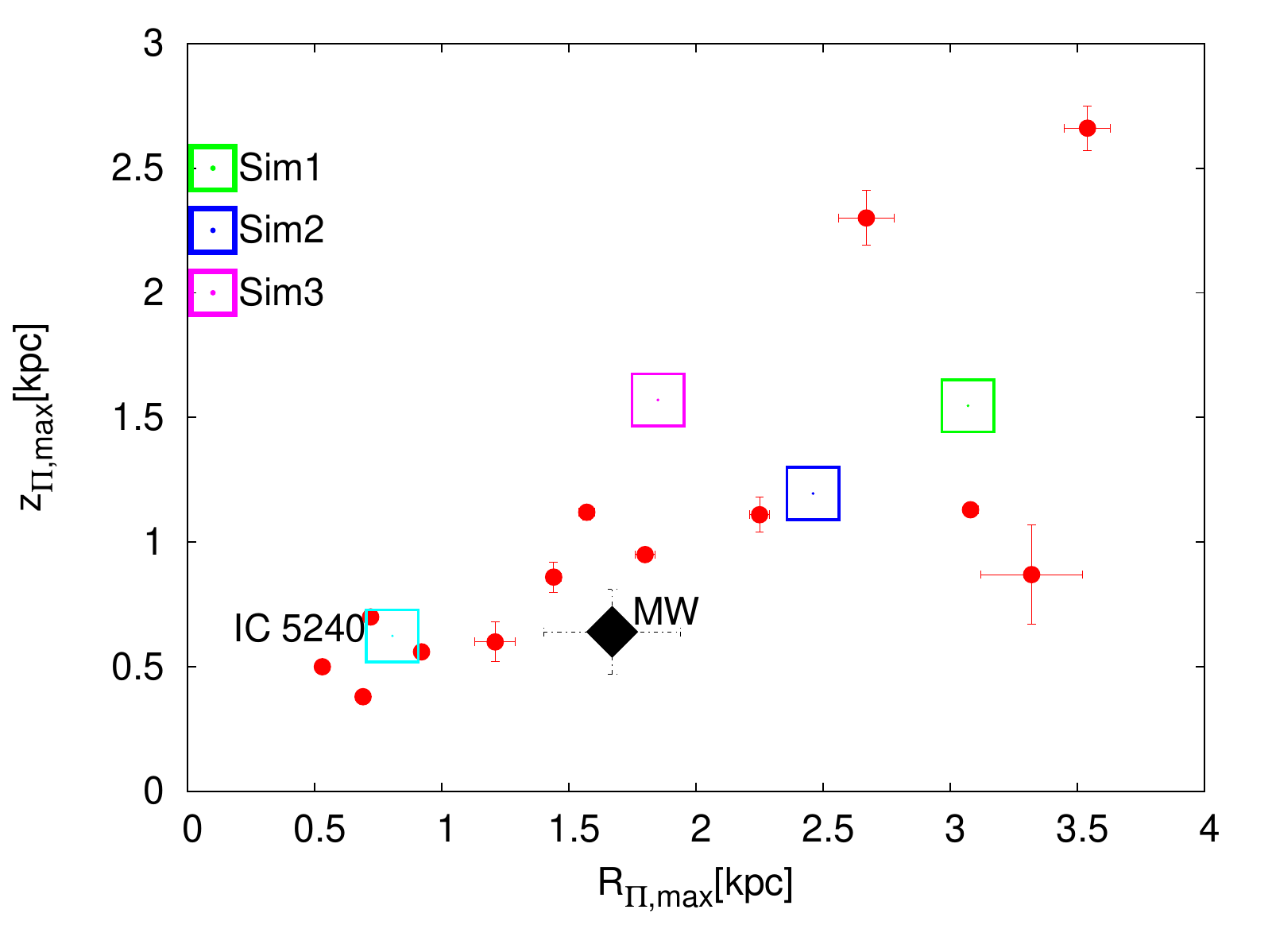}
\caption{Maximum perpendicular departure of peanuts above, and in, the disk
  plane is plotted against the radial length of the peanut for edge-on disk
  galaxies (red dots) taken from \cite{CG2016}, and the Milky Way (MW) taken
  from \cite{Ciamburetal2017}.  Over plotted are the face-on peanuts from our
  simulations and IC~5240.  For the face-on peanuts, we have used
  $xy_{\Pi,max}$ --- the mathematical equivalent to $z_{\Pi,max}$ (see
  footnote~\ref{foot1}).}
\label{fig:Rmax-zmax}
\end{figure}

\begin{figure}
\includegraphics[width=\columnwidth]{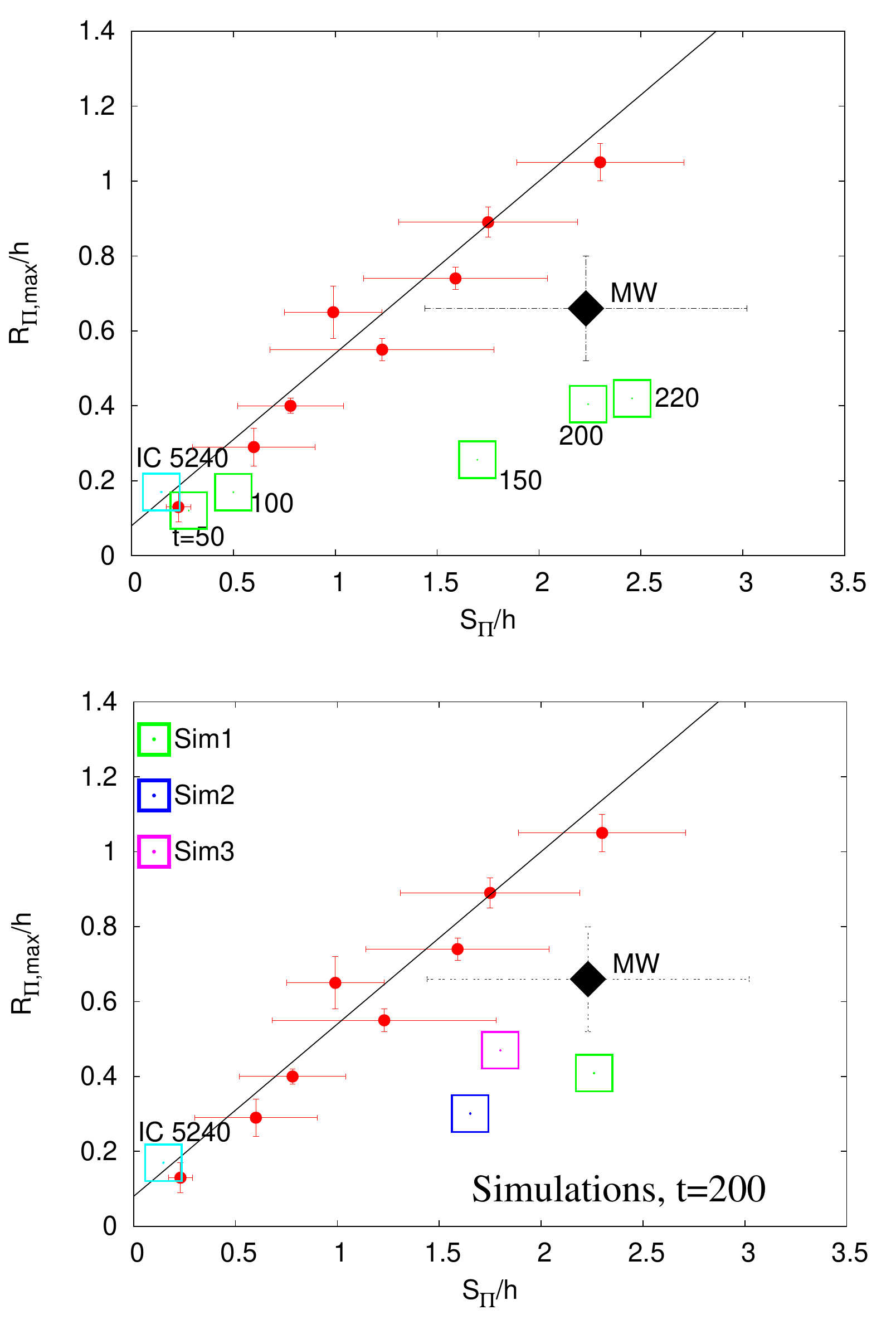}
\caption{Radial length versus integrated strength of the peanut structures
  seen in galaxies taken from \cite{CG2016} (red dots). Over plotted on this
  are our simulated face-on peanuts, IC 5240, and the Milky Way (MW,
  \cite{Ciamburetal2017}).  Solid line is the linear regression between
  $R_{\Pi}$ and $S_{\Pi}$, see \cite{CG2016}.  Top panel: For Sim1 at t=50,
  100, 150, 200, 220. Bottom panel: simulated peanuts at t=200 (12~Gyr).}
\label{fig:Rpi-Spi}
\end{figure}

Despite its disk inclination of $i=$ 49$^{\circ}$, we find similar features in
the $B_6$ profile and the surface brightness profile as to those seen in the
simulation. The top panel of Fig.~\ref{Decom-B6-IC5240} shows the
decomposition of the major-axis surface brightness profile of IC~5240, whose
constituents are found to be the same as in the simulations: a classical
S\`ersic bulge $n=1.2$, a double exponential disk, a Ferrers bar and a
Gaussian ring/ansae. This set of components worked well for all the systems
explored.

Similarly, the radial $B_6$ profile of IC~5240 (bottom panel of
Fig.~\ref{Decom-B6-IC5240}) exhibits a positive bump for the
peanut with its maximum around $\approx 15\arcsec$ ($\approx 2$~kpc). 
The bar in our model ends at a distance of $\approx 30\arcsec$ ($\approx 3.9$~kpc). 
The second positive bump in the $B_6$ profile plus the subsequent dip are due to 
the ansae/ring at the end of the bar and beyond.  The peanut parameters for IC~5240 are 
included in Table~\ref{tab:param2}.

\begin{figure*}
\rotatebox{0}{\includegraphics[width=1.9\columnwidth]{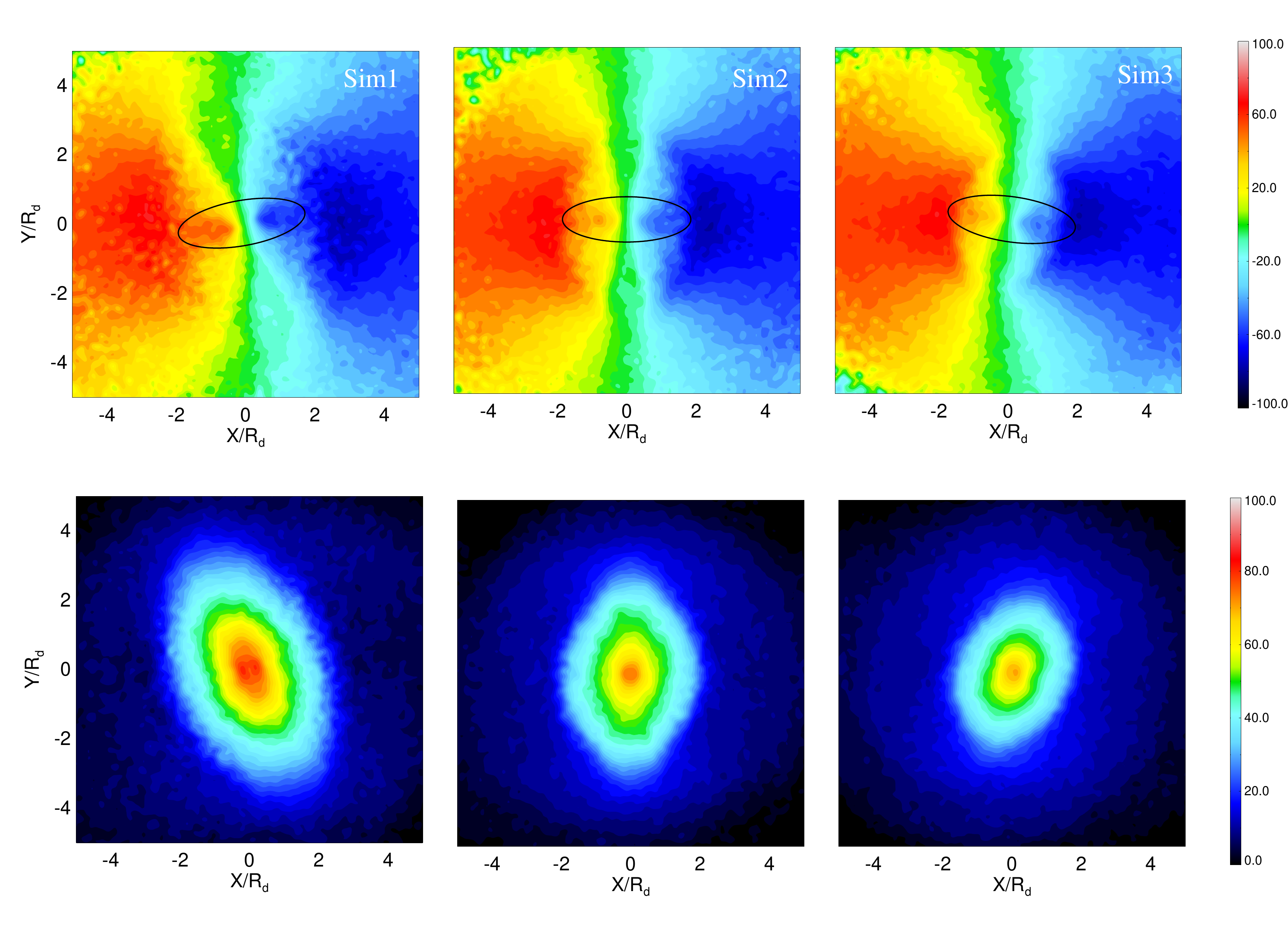}}
\caption{Top panel: Line-of-sight velocity maps for all three models, taken at
  $t=200$ (12 Gyr). 
The models are oriented such that the disk has an angle of $30^{\circ}$ from
face-on.  Bottom panel: Same as above but for the velocity dispersion. 
A ``kinematic pinch'' along the bar minor-axis is evident in the velocity map,
indicated by the black ellipses. The kinematic maps have the same color bar indicated on the right.}
\label{fig_kinemat}
\end{figure*}

\begin{figure}
\rotatebox{0}{\includegraphics[width=0.9\columnwidth]{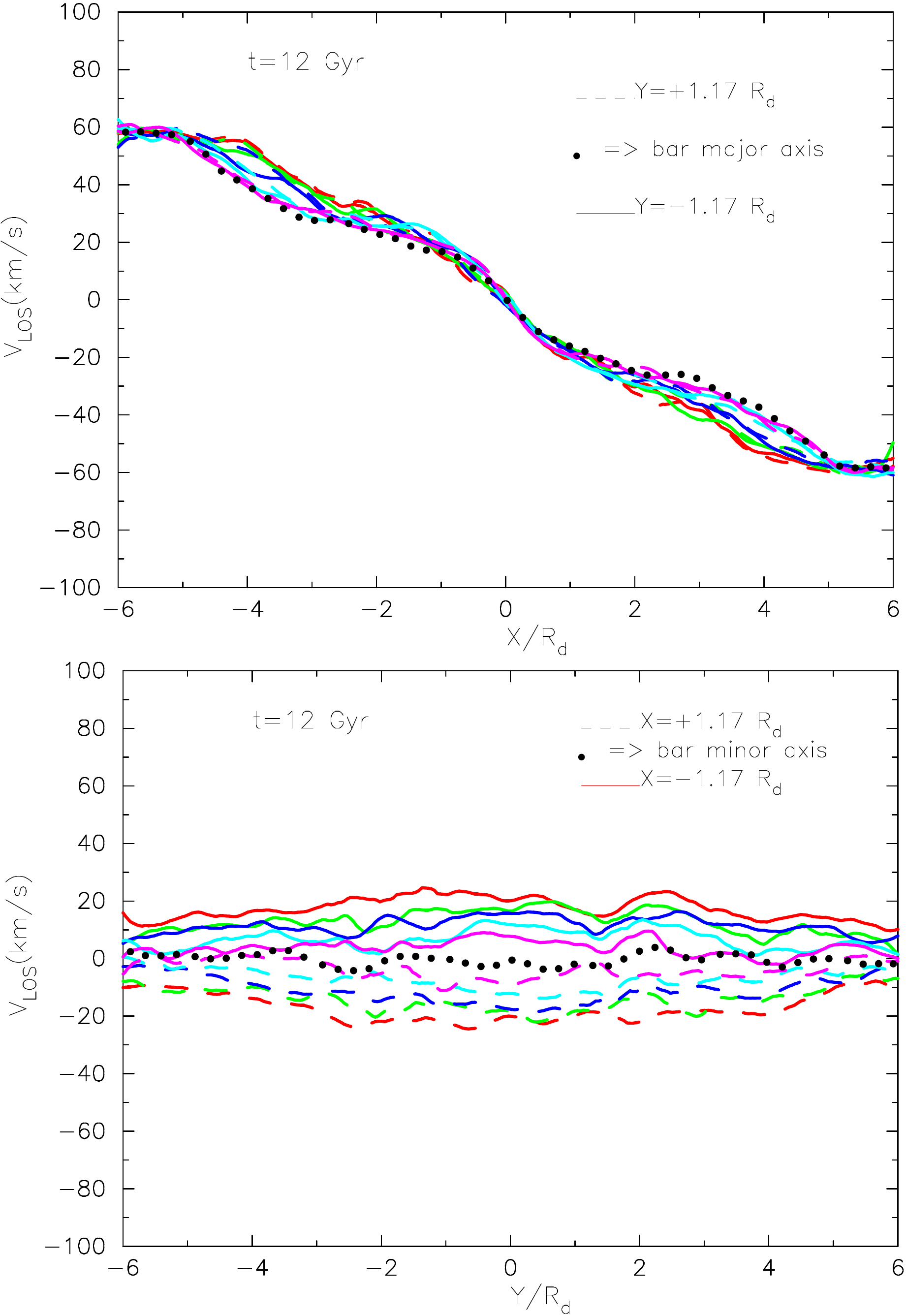}}
\caption{Top: Line-of-sight velocity profiles along slits parallel to the bar major axis
 at $t=200$ (12 Gyr) for Sim1. The model's disk is oriented at an angle of
 $30^{\circ}$ from face-on.  Five slits are placed above the bar
 major axis (positive Y-values) and five below. The furthest slits are marked
 in red (solid and dashed lines) at $Y=\pm 1.17$~$R_d$ (where $R_d$ is the
 initial $t=0$ disk scale-length equal to 3 kpc). 
 The distance between two consecutive slits is $0.24$ $R_d = 720$~pc. 
Bottom: Same as above but for the minor axis velocity profiles.}
\label{fig_velLOS}
\end{figure}

\begin{table}
\begin{center}
\caption{More peanut parameters}
\label{tab:param2}
\begin{tabular}{cccc}
\hline
                      &	Sim2   & Sim3   &  IC~5240\\
\hline
$\Pi_{max}$   	      & 0.073  & 0.055  &  0.014\\ 
\hline
$R_{\Pi,max}$ [kpc]   & 2.46   & 1.85   & 0.80 \\ 
$xy_{\Pi,max}$[kpc]	      & 1.19   & 1.57   & 0.62\\ 
$S_{\Pi}$[kpc]        & 13.5   & 7.07   & 0.70\\ 
$W_{\Pi}$[kpc]        & 2.40   & 1.82   & 0.67\\  
\hline
$h_{\rm disk}$[kpc]   & 8.17   & 3.93   & 4.8\\
\hline
\end{tabular}
\end{center}
 
Quantitative peanut parameters for the evolved ($t=200$, 12 Gyr) galaxy models
Sim2 and Sim3, and for the real galaxy IC~5240.  The quantity $h_{\rm disk}$
denotes the inner exponential scale length for the disk in Sim2, Sim3 and
IC~5240.
\end{table}

\section{Peanut scaling diagrams}
\label{sec:compare}

We use the new scaling relations for the peanut structures presented in
\cite{CG2016} to compare our simulated 3D peanuts. 

Fig.~\ref{fig:Rmax-zmax} shows the scaling diagram involving the radial length
of a peanut, $R_{\Pi,max}$, to its perpendicular departure from the
major-axis, $xy_{\Pi,max}$ (or $z_{\Pi,max}$).  Note, this denotes the isodensity
contour or isophote where the peanut appears strongest; it is not some fainter
contour towards the end of the bar, marking the full extent of the
peanut. Parameters for real galaxies, except IC~5240 are taken from
\cite{CG2016} and that of the Milky Way (MW) are from
\cite{Ciamburetal2017}. All three simulated peanuts follow the general trend shown
by the observed peanuts

In Fig.~\ref{fig:Rpi-Spi}, we compare the integrated strength of our simulated
peanuts with their radial extent, along with other galaxies including the MW.
For the peanut structures in galaxies from \cite{CG2016}, there seems to be a
linear relation between the strength and length of the peanut: the longer the
peanut, the stronger it is. After 12~Gyr, our simulated peanuts (and the Milky
Way) seem to deviate from this linear relation defined by the other galaxies.
This relation can perhaps now be understood as an age effect. From Sim1, we
learned that as the system evolved, the peanut (and bar) gets longer and
stronger (see Table~\ref{tab:param1}).  In passing, we note that unlike with
the simulations, within the sample of real galaxies, the bars are not always
viewed perpendicular to the line-of-sight.  Consequently, the projected
quantities $R_{\Pi,max}$ and $W_{\Pi}$, and thus $S_{\Pi}$, will be reduced
from their intrinsic value.

We additionally present the time evolution of the peanut that grows in
model Sim1. At $t=50$ (3~Gyr) and $t=100$ (6~Gyr), the peanut from Sim1 is close to the linear
regression followed by other galaxies and IC 5240.  As time progresses, the
Sim1 peanut gets stronger and eventually its strength becomes comparable to that
of the Milky Way's peanut. Overall, our simulated peanuts are stronger when
compared to peanut structures in other galaxies. 
This is perhaps not surprising given the relative strength of the peanut to
the underlying disk, as seen in Figure~\ref{fig:t200} and traced by $\Pi_{max}$. 
Off the major-axis, Sim1 has contributed much of its disk stars to the bar and
peanut (Figure~\ref{fig:t200}), this is less true in Sim3 which resides
closest to the line in Figure~\ref{fig:Rpi-Spi}.  One of the uncertainties that
may affect such a comparison is the orientation of the bar/peanut in the plane 
of the disk (as mentioned above). Measurements from our simulations are on the
higher side (in fact, they are the maximum) as the bar is oriented perpendicular to the 
line-of-sight. Any deviation from this, would lead to a decrease in the measured
values of the parameters defining the peanut structures. This would apply to all
the scaling relations that we use to compare with our simulated peanuts. 

In the following section, we discuss the kinematic scaling relations followed
by our simulated peanuts.

\section{Kinematics of face-on peanut structures}
\label{sec:bkin}

\subsection{Kinematic pinches}

Alongside photometry, the kinematics of galaxies have also played an important
role, not only in characterizing boxy/peanut/X-shapes, but also enhancing our
understanding of their basic physical nature. A typical such peanut structure
would exhibit cylindrical rotation in edge-on projection (where the rotation
axis, $z$, is perpendicular to the line-of-sight). In that case, the
line-of-sight velocity at a given projected radius becomes independent of z, i.e.,
\ $d V_{LOS}(X=X_0, z)/dz \simeq 0$, where $X=X_0$ is any location within the peanut 
structure along the major axis. The cylindrical rotation is, therefore, a standard kinematic
proxy for a peanut structure, although there are exceptions both in
observations and simulated peanut structures \citep{Williamsetal2011,SahaGerhard2013}.

In Fig.~\ref{fig_kinemat}, we show the 2D line-of-sight velocity and velocity
dispersion maps ($i_{\rm disk}=30^{\circ}$) for all three models at $t=12$
Gyr; see Fig.~\ref{fig_contour} for their respective surface density maps. The
velocity maps clearly show rotating model galaxies, but their inner parts are
rather slowly rotating (although they were not initially).  At t=12 Gyr,
$V_{max}/\sigma_{in} \simeq 1.5$ for Sim1; 1.7 and 1.9 for Sim2 and Sim3
respectively. Interestingly, these face-on peanuts are associated with a
``Kinematic Pinch'' along the bar minor axis. We have already identified such
a pinch in the surface density maps. The pinch (both photometric and
kinematic) is stronger in Sim1, and diminishes gradually in Sim3. When the
kinematic maps are available for IC~5240, we would expect to observe such a
pinch.

\subsection{Cylindrical rotation?}
In the following, we examine the major and minor axis velocity profiles of the stars
in near face-on projection. 

The extent of the 3D peanut is about 1~$R_d$, whereas the outer X-shape 
extends to more than 2~$R_d$ in edge-on projection, see Fig.~\ref{fig:t200}.
The upper panel of Fig.~\ref{fig_velLOS} shows the LOS velocity profiles along the major of the bar 
at $t=12$~Gyr for $11$ slits placed parallel to the bar major axis. The outer slits are at
$Y = \pm 1.17 R_d$, basically covering most of the bar along Y-axis. There is
a clear wavy nature in the profiles. We have checked this also at earlier times.
At $t=4$~Gyr, when the bar still shows no pinching but has formed a boxy structure, 
the velocity profile shows small scale wavy signature. At later times, such wavy features 
in the velocity profile become much more prominent. 

Apart from the wavy feature, we notice that within about $X = 2$~$R_d$, the
velocity profiles are hardly separable, that is, the velocity
profiles along the bar minor axis change very little: $dV_{LOS}({X<2R_d},
Y)/dY \simeq 0$.  This is exactly analogous to the cylindrical rotation
observed in edge-on projection - in other words, these face-on simulated peanuts
exhibit cylindrical rotation even when the disk is oriented in such low inclination angle.

In the bottom panel of Fig.~\ref{fig_velLOS}, the minor axis velocity profiles are displayed
along 11 slits placed perpendicular to the bar major axis. The outer most slits are at
$X=1.17 R_d$ on either side of the bar minor axis. Obviously, the variation of $V_{LOS}$ with 
respect to Y is insignificant. It turns out that the inner part of the peanut structure
(approximately the zoom-in area shown in Fig.~\ref{fig:t200}) exhibit
strong cylindrical rotation from high inclination angle ($i=90^{\circ}$) to as low as $30^{\circ}$.  
It turns out that at $t=12$~Gyr, the bar pattern speed is $\Omega_{B} \simeq
6$ km s$^{-1}$ kpc$^{-1}$, a value very close to the slope $dV_{LOS}/dX$ within the inner $\sim 1.5~R_d$.
Note that all three bars in our simulations showing face-on peanut are indeed slow bars i.e., 
their corotation to bar length is greater than 1.4.

\begin{figure}
\includegraphics[width=1.0\columnwidth]{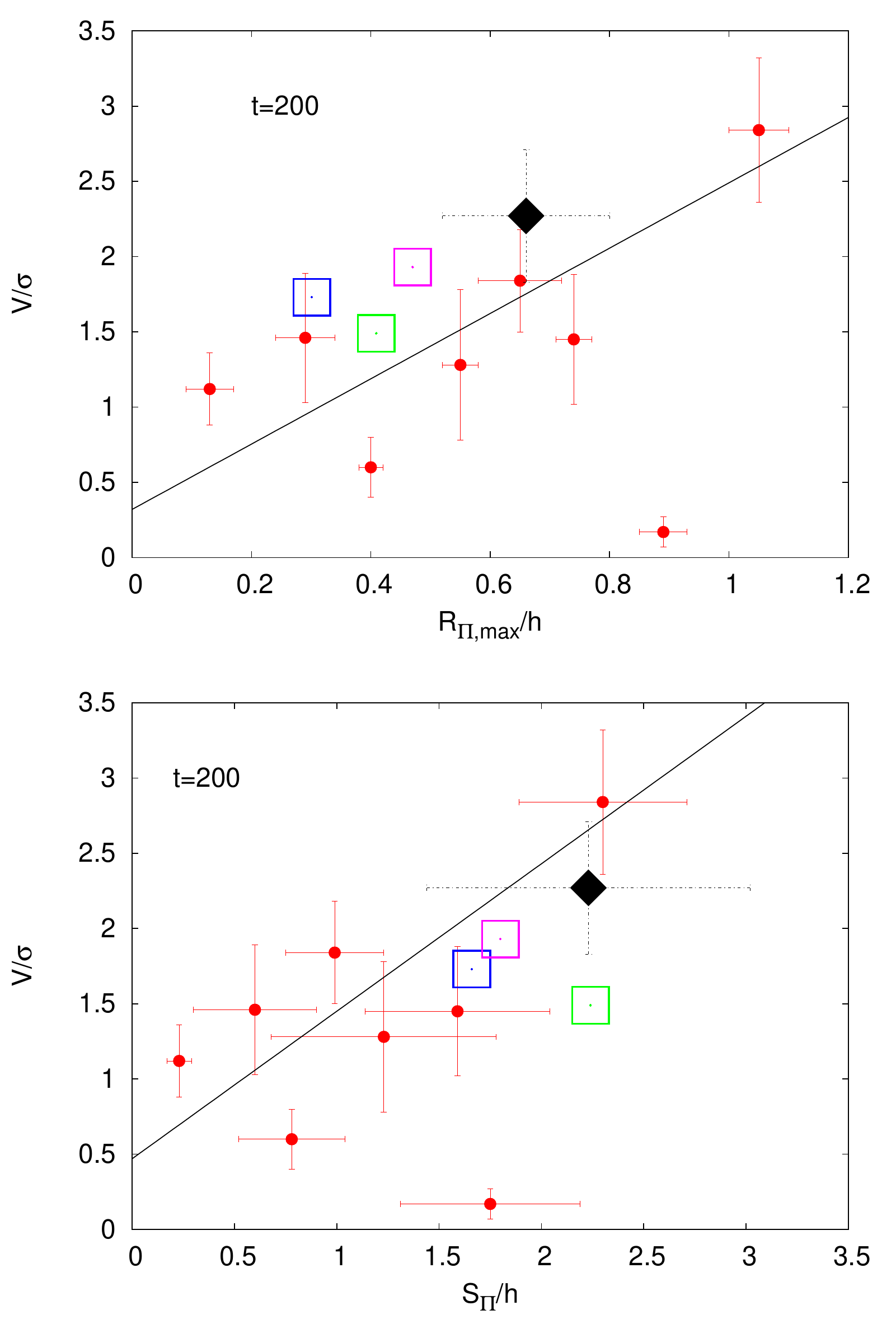}
\caption{The dependence of $V/\sigma$ on the length and strength of the peanut
  structures in a sample of edge-on galaxies \citep{CG2016} (red dots) and the
  MW \citep{Ciamburetal2017} compared with the face-on peanuts in our
  simulated models.  The symbols have the same meaning as in
  Fig.~\ref{fig:Rmax-zmax}.}
\label{fig:vsig-RpiSpi}
\end{figure}

\subsection{Kinematic scaling diagram}

In Fig.~\ref{fig:vsig-RpiSpi}, we present a kinematic scaling diagram for our
simulated peanuts and compare with those in \cite{CG2016}. We computed
$V/\sigma$ for each of the simulated models at different epochs during the
evolution; such that $V$ is the maximum rotation velocity in the disk and
$\sigma$ is the central velocity dispersion (averaged within a 1.5 kpc region
about the center).  The quantity $V/\sigma$ is indicative of how slowly or
fast the galaxy (not the peanut) is rotating. Sim1, which formed the strongest
bar and strongest peanut, is the slowest rotating galaxy among the models
presented here. Fig.~\ref{fig:vsig-RpiSpi} shows the variation of $V/\sigma$
with the length and strength of the simulated peanuts and compared with Milky
Way peanut.  The simulated peanuts are in broad agreement with the observed
peanut kinematic scaling relation. It would be instructive to compare these
with a bigger peanut galaxy sample in a future study, including more galaxies
with face-on disks.

\section{Discussion and Conclusions}
\label{sec:discussion}

We have addressed an important issue on the physical nature of boxy/peanut
structures in local disk galaxies. The conventional view is that these
structures form as a result of the vertical buckling instability of a
bar \citep{CombesSanders1981}. During this instability, stars are excited at 
the vertical inner Lindblad
resonances \citep{Pfenniger1985, PfennigerFriedli1991,MV2006}, resulting in a
vertically puffed up bar with a characteristic peanut morphology, most
prominent when the host galaxy's disk is viewed in an edge-on projection. This
special viewing angle restriction had resulted in a low number of galaxies
observed with these peanuts, although recently more and more disk galaxies are
being reported to host such peanuts as near-(face-on) diagnostics are being
identified \citep{Mendez-Abreuetal2008, IannuzziAthanassoula2015,
  ErwinDebattista2016,ErwinDebattista2017}. In addition to the viewing angle
restriction, there are other structural components in a disk galaxy that are known 
to contribute to the weakening of the bar and peanut structures intrinsically - they are mainly
the gas component \citep{Berentzenetal1998,Athanassoulaetal2013} and the central mass 
concentrations \citep{Hasanetal1993,Bournaudetal2005}. Despite these, a number of observed galaxies
have shown peanut signatures in face-on projection \citep{Laurikainenetal2011,ErwinDebattista2013}. So
the actual reason why some galaxies reveal peanuts in face-on, remains to investigated. Our 
simulatins are purely collisionless, and hence are unable to comment on the imapct of gas on the 
final appearance of the peanut structure. However, our simulations show 
it clearly that as the peanut structure in edge-on projection gets stronger, its face-on counterpart 
also becomes prominent (see, Sim1 evolution above).

In our simulations of cold stellar disks with low bulge-to-disk mass ratios, 
bars with strong three-dimensional peanut-shaped structures form.
The results from our simulations show that there is a strong pinch
along the bar-minor axis, giving it a peanut shape in the X-Y plane of the
galaxy disk and this is {\em not} due to any projection effect, see
Fig.~\ref{fig:t200} and Fig.~\ref{fig_contour}. As expected, the pinch in the
density is also accompanied by its signature in the kinematic map, see
Fig.~\ref{fig_kinemat}. Such a density-kinematic pinch is highly prominent in
Sim1 and least in Sim3. Recently, \cite{PatsisKatsanikas2014} has reported
possible orbital connection between the face-on and edge-on peanut
morphology. However, the exact physical conditions that lead to the formation
of these 3D X/peanut structures need further investigation.

It is important to distinguish our simulated face-on peanuts from
barlenses. Barlenses are thought to have a dynamical connection with vertical ($z$-direction)
boxy/peanut structures: as shown by \cite{Athanassoulaetal2015}, barlenses and
boxy/peanut structures are essentially the same feature viewed at a different
angle; further evidence is also found in observations 
\citep{Herrera-Endoquietal2017}. In a similar fashion, barlenses and X-shaped
features are also being compared by \cite{LaurikainenSalo2017}. All three of
our simulated models of galaxies host boxy/peanut/X-shaped structures when their
disks are viewed edge-on {\it and} face-on. This is not due to any projection effect; 
there is a peanut both in edge-on and face-on projection - making this a truly 
three-dimensional peanut. In addition to the well-known cylindrical rotation when 
viewed edge-on, we found the signature of cylindrical rotation to extend even in low inclination
angle, i.e., in near face-on projection. We have shown clear cylindrical
rotation in all the simulated 3D peanuts at $30^{\circ}$ inclination from
face-on.

Our primary conclusions from this work are:

1. Our simulations have revealed strong peanut structures associated with an ansae 
and an outer ring at the end of the bar --- as seen in the galaxy IC~5240.

2.  There is a close morphological resemblance between our simulations and IC~5240.  
The structural components (a classical S\'ersic bulge with $n\approx 1$, a double 
exponential disk, a Ferrers bar, and a Gaussian ring/ansae) that were required to 
model the images of our three simulated galaxies and IC~5240 are the same, albeit 
with different parameter values. The broader ring in Sim2 required an additional 
Gaussian ring component.

3. Beyond the cutoff of the Peanut/X/bowtie structures, which have their maximum 
strength (relative to the galaxy light) at ~0.5 times the length of the bar, and 
extend out towards the end of the bar, we discover a positive/negative feature in 
the $B_6$ profile that is associated with the ansae/ring when the disk is viewed 
face-on.

4. The peanut-shaped structure appears to contribute little signal to the major-axis light profile; 
the decomposition of this profile does not require a peanut component in either the simulations 
or the real galaxy. We therefore consider it likely that some past works claiming to have found 
and measured the peanut-shaped structure arising from the bar, i.e. the ill-named "pseudobulge", 
via the decomposition of major-axis light profiles, have in fact not measured the galaxy's 
"pseudobulge", but have instead detected and measured the classical bulge.

5. We have found that in various structural and kinematic scaling diagrams 
(Figures~\ref{fig:vsig-RpiSpi}, \ref{fig:Rpi-Spi} \ref{fig:vsig-RpiSpi}), the quantitative 
peanut parameters measured in our face-on simulated galaxies are in broad agreement with 
quantitative peanut parameters measured in real, relatively edge-on, galaxies.

6. With a near-face-on projection, there is a {\it kinematic pinch} along the bar
minor axis --- coinciding with the density pinch. 
Prominent cylindrical rotation is evident even in near face-on projection.

\acknowledgements
A.W.G. was supported under the Australian Research Council's funding scheme (DP17012923).
We would like to thank the anonymous referee for useful comments. We would like to 
thank Ron Buta for pointing out IC 5240 as an example of a possible face-on peanut. All the 
simulations were performed at the IUCAA HPC centre.


\begin{thebibliography}{72}
\expandafter\ifx\csname natexlab\endcsname\relax\def\natexlab#1{#1}\fi

\bibitem[{{Athanassoula}(2005)}]{Athanassoula2005}
{Athanassoula}, E. 2005, \mnras, 358, 1477

\bibitem[{{Athanassoula}(2016)}]{Athanassoula2016}
---. 2016, Galactic Bulges, 418, 391

\bibitem[{{Athanassoula} \& {Beaton}(2006)}]{AthanassoulaBeaton2006}
{Athanassoula}, E., \& {Beaton}, R.~L. 2006, \mnras, 370, 1499

\bibitem[{{Athanassoula} {et~al.}(2015){Athanassoula}, {Laurikainen}, {Salo},
  \& {Bosma}}]{Athanassoulaetal2015}
{Athanassoula}, E., {Laurikainen}, E., {Salo}, H., \& {Bosma}, A. 2015, \mnras,
  454, 3843

\bibitem[{{Athanassoula} {et~al.}(2013){Athanassoula}, {Machado}, \&
  {Rodionov}}]{Athanassoulaetal2013}
{Athanassoula}, E., {Machado}, R.~E.~G., \& {Rodionov}, S.~A. 2013, \mnras,
  429, 1949

\bibitem[{{Athanassoula} \& {Misiriotis}(2002)}]{Athanamisi2002}
{Athanassoula}, E., \& {Misiriotis}, A. 2002, \mnras, 330, 35

\bibitem[{{Berentzen} {et~al.}(1998){Berentzen}, {Heller}, {Shlosman}, \&
  {Fricke}}]{Berentzenetal1998}
{Berentzen}, I., {Heller}, C.~H., {Shlosman}, I., \& {Fricke}, K.~J. 1998,
  \mnras, 300, 49

\bibitem[{{Binney} \& {Tremaine}(1987)}]{BT1987}
{Binney}, J., \& {Tremaine}, S. 1987, {Galactic dynamics}, ed. {Binney, J.~\&
  Tremaine, S.}

\bibitem[{{Block} {et~al.}(2001){Block}, {Puerari}, {Knapen}, {Elmegreen},
  {Buta}, {Stedman}, \& {Elmegreen}}]{Blocketal2001}
{Block}, D.~L., {Puerari}, I., {Knapen}, J.~H., {et~al.} 2001, \aap, 375, 761

\bibitem[{{Bournaud} {et~al.}(2005){Bournaud}, {Combes}, \&
  {Semelin}}]{Bournaudetal2005}
{Bournaud}, F., {Combes}, F., \& {Semelin}, B. 2005, \mnras, 364, L18

\bibitem[{{Bureau} \& {Freeman}(1999)}]{BureauFreeman1999}
{Bureau}, M., \& {Freeman}, K.~C. 1999, \aj, 118, 126

\bibitem[{{Buta}(1995)}]{Buta1995}
{Buta}, R. 1995, \apjs, 96, 39

\bibitem[{{Buta} \& {Crocker}(1991)}]{ButaCrocker1991}
{Buta}, R., \& {Crocker}, D.~A. 1991, \aj, 102, 1715

\bibitem[{{Buta} {et~al.}(2010){Buta}, {Sheth}, {Regan}, {Hinz}, {Gil de Paz},
  {Men{\'e}ndez-Delmestre}, {Munoz-Mateos}, {Seibert}, {Laurikainen}, {Salo},
  {Gadotti}, {Athanassoula}, {Bosma}, {Knapen}, {Ho}, {Madore}, {Elmegreen},
  {Masters}, {Comer{\'o}n}, {Aravena}, \& {Kim}}]{Butaetal2010}
{Buta}, R.~J., {Sheth}, K., {Regan}, M., {et~al.} 2010, \apjs, 190, 147

\bibitem[{{Chung} \& {Bureau}(2004)}]{ChungBureau2004}
{Chung}, A., \& {Bureau}, M. 2004, \aj, 127, 3192

\bibitem[{{Ciambur}(2015)}]{C15Isofit}
{Ciambur}, B.~C. 2015, \apj, 810, 120

\bibitem[{{Ciambur}(2016)}]{Ciambur2016}
---. 2016, \pasa, 33, e062

\bibitem[{{Ciambur} \& {Graham}(2016)}]{CG2016}
{Ciambur}, B.~C., \& {Graham}, A.~W. 2016, \mnras, 459, 1276

\bibitem[{{Ciambur} {et~al.}(2017){Ciambur}, {Graham}, \&
  {Bland-Hawthorn}}]{Ciamburetal2017}
{Ciambur}, B.~C., {Graham}, A.~W., \& {Bland-Hawthorn}, J. 2017, ArXiv e-prints

\bibitem[{{Combes}(2011)}]{Combes2011}
{Combes}, F. 2011, in IAU Symposium, Vol. 271, Astrophysical Dynamics: From
  Stars to Galaxies, ed. N.~H. {Brummell}, A.~S. {Brun}, M.~S. {Miesch}, \&
  Y.~{Ponty}, 119--126

\bibitem[{{Combes}(2016)}]{Combes2016}
{Combes}, F. 2016, Galactic Bulges, 418, 413

\bibitem[{{Combes} {et~al.}(1990){Combes}, {Debbasch}, {Friedli}, \&
  {Pfenniger}}]{Combesetal1990}
{Combes}, F., {Debbasch}, F., {Friedli}, D., \& {Pfenniger}, D. 1990, \aap,
  233, 82

\bibitem[{{Combes} \& {Sanders}(1981)}]{CombesSanders1981}
{Combes}, F., \& {Sanders}, R.~H. 1981, \aap, 96, 164

\bibitem[{{de Vaucouleurs} {et~al.}(1991){de Vaucouleurs}, {de Vaucouleurs},
  {Corwin}, {Buta}, {Paturel}, \& {Fouqu{\'e}}}]{deVaucouleurs1991}
{de Vaucouleurs}, G., {de Vaucouleurs}, A., {Corwin}, Jr., H.~G., {et~al.}
  1991, {Third Reference Catalogue of Bright Galaxies. Volume I: Explanations
  and references. Volume II: Data for galaxies between 0$^{h}$ and 12$^{h}$.
  Volume III: Data for galaxies between 12$^{h}$ and 24$^{h}$.}

\bibitem[{{Debattista} {et~al.}(2005){Debattista}, {Carollo}, {Mayer}, \&
  {Moore}}]{Debattistaetal2005}
{Debattista}, V.~P., {Carollo}, C.~M., {Mayer}, L., \& {Moore}, B. 2005, \apj,
  628, 678

\bibitem[{{Debattista} {et~al.}(2006){Debattista}, {Mayer}, {Carollo}, {Moore},
  {Wadsley}, \& {Quinn}}]{Debattistaetal2006}
{Debattista}, V.~P., {Mayer}, L., {Carollo}, C.~M., {et~al.} 2006, \apj, 645,
  209

\bibitem[{{Dwek} {et~al.}(1995){Dwek}, {Arendt}, {Hauser}, {Kelsall}, {Lisse},
  {Moseley}, {Silverberg}, {Sodroski}, \& {Weiland}}]{Dweketal1995}
{Dwek}, E., {Arendt}, R.~G., {Hauser}, M.~G., {et~al.} 1995, \apj, 445, 716

\bibitem[{{Elmegreen} {et~al.}(2008){Elmegreen}, {Bournaud}, \&
  {Elmegreen}}]{Elmegreenetal2008}
{Elmegreen}, B.~G., {Bournaud}, F., \& {Elmegreen}, D.~M. 2008, \apj, 688, 67

\bibitem[{{Erwin} \& {Debattista}(2013)}]{ErwinDebattista2013}
{Erwin}, P., \& {Debattista}, V.~P. 2013, \mnras, 431, 3060

\bibitem[{{Erwin} \& {Debattista}(2016)}]{ErwinDebattista2016}
---. 2016, \apjl, 825, L30

\bibitem[{{Erwin} \& {Debattista}(2017)}]{ErwinDebattista2017}
---. 2017, \mnras, 468, 2058

\bibitem[{{Evans}(1993)}]{Evans1993}
{Evans}, N.~W. 1993, \mnras, 260, 191

\bibitem[{{Graham}(2015)}]{GrahamConf2015}
{Graham}, A. 2015, Highlights of Astronomy, 16, 360

\bibitem[{{Graham}(2014)}]{GrahamConf2014}
{Graham}, A.~W. 2014, in Astronomical Society of the Pacific Conference Series,
  Vol. 480, Structure and Dynamics of Disk Galaxies, ed. M.~S. {Seigar} \&
  P.~{Treuthardt}, 185

\bibitem[{{Hasan} {et~al.}(1993){Hasan}, {Pfenniger}, \&
  {Norman}}]{Hasanetal1993}
{Hasan}, H., {Pfenniger}, D., \& {Norman}, C. 1993, \apj, 409, 91

\bibitem[{{Herrera-Endoqui} {et~al.}(2017){Herrera-Endoqui}, {Salo},
  {Laurikainen}, \& {Knapen}}]{Herrera-Endoquietal2017}
{Herrera-Endoqui}, M., {Salo}, H., {Laurikainen}, E., \& {Knapen}, J.~H. 2017,
  \aap, 599, A43

\bibitem[{{Iannuzzi} \& {Athanassoula}(2015)}]{IannuzziAthanassoula2015}
{Iannuzzi}, F., \& {Athanassoula}, E. 2015, \mnras, 450, 2514

\bibitem[{{Jogee} {et~al.}(2005){Jogee}, {Scoville}, \&
  {Kenney}}]{Jogeeetal2005}
{Jogee}, S., {Scoville}, N., \& {Kenney}, J.~D.~P. 2005, \apj, 630, 837

\bibitem[{{King}(1966)}]{King1966}
{King}, I.~R. 1966, \aj, 71, 64

\bibitem[{{Kuijken} \& {Dubinski}(1995)}]{KD1995}
{Kuijken}, K., \& {Dubinski}, J. 1995, \mnras, 277, 1341

\bibitem[{{Kuijken} \& {Merrifield}(1995)}]{KuijkenMerrifield1995}
{Kuijken}, K., \& {Merrifield}, M.~R. 1995, \apjl, 443, L13

\bibitem[{{Laurikainen} \& {Salo}(2017)}]{LaurikainenSalo2017}
{Laurikainen}, E., \& {Salo}, H. 2017, \aap, 598, A10

\bibitem[{{Laurikainen} {et~al.}(2014){Laurikainen}, {Salo}, {Athanassoula},
  {Bosma}, \& {Herrera-Endoqui}}]{Laurikainenetal2014}
{Laurikainen}, E., {Salo}, H., {Athanassoula}, E., {Bosma}, A., \&
  {Herrera-Endoqui}, M. 2014, \mnras, 444, L80

\bibitem[{{Laurikainen} {et~al.}(2011){Laurikainen}, {Salo}, {Buta}, \&
  {Knapen}}]{Laurikainenetal2011}
{Laurikainen}, E., {Salo}, H., {Buta}, R., \& {Knapen}, J.~H. 2011, \mnras,
  418, 1452

\bibitem[{{Laurikainen} {et~al.}(2004){Laurikainen}, {Salo}, {Buta}, \&
  {Vasylyev}}]{Laurikainenetal2004}
{Laurikainen}, E., {Salo}, H., {Buta}, R., \& {Vasylyev}, S. 2004, \mnras, 355,
  1251

\bibitem[{{L{\"u}tticke} {et~al.}(2000{\natexlab{a}}){L{\"u}tticke}, {Dettmar},
  \& {Pohlen}}]{Luttickeetal2000a}
{L{\"u}tticke}, R., {Dettmar}, R.-J., \& {Pohlen}, M. 2000{\natexlab{a}},
  \aaps, 145, 405

\bibitem[{{L{\"u}tticke} {et~al.}(2000{\natexlab{b}}){L{\"u}tticke}, {Dettmar},
  \& {Pohlen}}]{Luttickeetal2000b}
---. 2000{\natexlab{b}}, \aap, 362, 435

\bibitem[{{Martinez-Valpuesta} {et~al.}(2007){Martinez-Valpuesta}, {Knapen}, \&
  {Buta}}]{MVKB07}
{Martinez-Valpuesta}, I., {Knapen}, J.~H., \& {Buta}, R. 2007, \aj, 134, 1863

\bibitem[{{Martinez-Valpuesta} \& {Shlosman}(2004)}]{MV2004}
{Martinez-Valpuesta}, I., \& {Shlosman}, I. 2004, \apjl, 613, L29

\bibitem[{{Martinez-Valpuesta} {et~al.}(2006){Martinez-Valpuesta}, {Shlosman},
  \& {Heller}}]{MV2006}
{Martinez-Valpuesta}, I., {Shlosman}, I., \& {Heller}, C. 2006, \apj, 637, 214

\bibitem[{{McMillan} \& {Dehnen}(2007)}]{McMillan2007}
{McMillan}, P.~J., \& {Dehnen}, W. 2007, \mnras, 378, 541

\bibitem[{{M{\'e}ndez-Abreu} {et~al.}(2008){M{\'e}ndez-Abreu}, {Corsini},
  {Debattista}, {De Rijcke}, {Aguerri}, \& {Pizzella}}]{Mendez-Abreuetal2008}
{M{\'e}ndez-Abreu}, J., {Corsini}, E.~M., {Debattista}, V.~P., {et~al.} 2008,
  \apjl, 679, L73

\bibitem[{{Mould} {et~al.}(2000){Mould}, {Huchra}, {Freedman}, {Kennicutt},
  {Ferrarese}, {Ford}, {Gibson}, {Graham}, {Hughes}, {Illingworth}, {Kelson},
  {Macri}, {Madore}, {Sakai}, {Sebo}, {Silbermann}, \&
  {Stetson}}]{Mouldetal2000}
{Mould}, J.~R., {Huchra}, J.~P., {Freedman}, W.~L., {et~al.} 2000, \apj, 529,
  786

\bibitem[{{O'Neill} \& {Dubinski}(2003)}]{ONeilDubinski2003}
{O'Neill}, J.~K., \& {Dubinski}, J. 2003, \mnras, 346, 251

\bibitem[{{Patsis} \& {Katsanikas}(2014)}]{PatsisKatsanikas2014}
{Patsis}, P.~A., \& {Katsanikas}, M. 2014, \mnras, 445, 3546

\bibitem[{{Patsis} {et~al.}(2002){Patsis}, {Skokos}, \&
  {Athanassoula}}]{Patsisetal2002}
{Patsis}, P.~A., {Skokos}, C., \& {Athanassoula}, E. 2002, \mnras, 337, 578

\bibitem[{{Pfenniger}(1985)}]{Pfenniger1985}
{Pfenniger}, D. 1985, \aap, 150, 112

\bibitem[{{Pfenniger} \& {Friedli}(1991)}]{PfennigerFriedli1991}
{Pfenniger}, D., \& {Friedli}, D. 1991, \aap, 252, 75

\bibitem[{{Pfenniger} \& {Norman}(1990)}]{PfennigerNorman1990}
{Pfenniger}, D., \& {Norman}, C. 1990, \apj, 363, 391

\bibitem[{{Planck Collaboration} {et~al.}(2016){Planck Collaboration}, {Ade},
  {Aghanim}, {Arnaud}, {Ashdown}, {Aumont}, {Baccigalupi}, {Banday},
  {Barreiro}, {Bartlett}, \& et~al.}]{Planck2016}
{Planck Collaboration}, {Ade}, P.~A.~R., {Aghanim}, N., {et~al.} 2016, \aap,
  594, A13

\bibitem[{{Quillen} {et~al.}(1997){Quillen}, {Kuchinski}, {Frogel}, \&
  {DePoy}}]{Quillenetal1997}
{Quillen}, A.~C., {Kuchinski}, L.~E., {Frogel}, J.~A., \& {DePoy}, D.~L. 1997,
  \apj, 481, 179

\bibitem[{{Raha} {et~al.}(1991){Raha}, {Sellwood}, {James}, \&
  {Kahn}}]{Rahaetal1991}
{Raha}, N., {Sellwood}, J.~A., {James}, R.~A., \& {Kahn}, F.~D. 1991, \nat,
  352, 411

\bibitem[{{Saha} \& {Gerhard}(2013)}]{SahaGerhard2013}
{Saha}, K., \& {Gerhard}, O. 2013, \mnras, 430, 2039

\bibitem[{{Saha} {et~al.}(2012){Saha}, {Martinez-Valpuesta}, \&
  {Gerhard}}]{Sahaetal2012}
{Saha}, K., {Martinez-Valpuesta}, I., \& {Gerhard}, O. 2012, \mnras, 421, 333

\bibitem[{{Saha} {et~al.}(2013){Saha}, {Pfenniger}, \& {Taam}}]{Sahaetal2013}
{Saha}, K., {Pfenniger}, D., \& {Taam}, R.~E. 2013, \apj, 764, 123

\bibitem[{{Saha} {et~al.}(2010){Saha}, {Tseng}, \& {Taam}}]{Sahaetal2010}
{Saha}, K., {Tseng}, Y., \& {Taam}, R.~E. 2010, \apj, 721, 1878

\bibitem[{{Sellwood} \& {Carlberg}(2014)}]{SellwoodCarlberg2014}
{Sellwood}, J.~A., \& {Carlberg}, R.~G. 2014, \apj, 785, 137

\bibitem[{{Sheth} {et~al.}(2010){Sheth}, {Regan}, {Hinz}, {Gil de Paz},
  {Men{\'e}ndez-Delmestre}, {Mu{\~n}oz-Mateos}, {Seibert}, {Kim},
  {Laurikainen}, {Salo}, {Gadotti}, {Laine}, {Mizusawa}, {Armus},
  {Athanassoula}, {Bosma}, {Buta}, {Capak}, {Jarrett}, {Elmegreen},
  {Elmegreen}, {Knapen}, {Koda}, {Helou}, {Ho}, {Madore}, {Masters},
  {Mobasher}, {Ogle}, {Peng}, {Schinnerer}, {Surace}, {Zaritsky},
  {Comer{\'o}n}, {de Swardt}, {Meidt}, {Kasliwal}, \&
  {Aravena}}]{Shethetal2010}
{Sheth}, K., {Regan}, M., {Hinz}, J.~L., {et~al.} 2010, \pasp, 122, 1397

\bibitem[{{Springel} {et~al.}(2001){Springel}, {Yoshida}, \&
  {White}}]{Springeletal2001}
{Springel}, V., {Yoshida}, N., \& {White}, S.~D.~M. 2001, \na, 6, 79

\bibitem[{{Toomre}(1969)}]{Toomre1969}
{Toomre}, A. 1969, \apj, 158, 899

\bibitem[{{Williams} {et~al.}(2011){Williams}, {Zamojski}, {Bureau},
  {Kuntschner}, {Merrifield}, {de Zeeuw}, \& {Kuijken}}]{Williamsetal2011}
{Williams}, M.~J., {Zamojski}, M.~A., {Bureau}, M., {et~al.} 2011, \mnras, 414,
  2163

\bibitem[{{Yoshino} \& {Yamauchi}(2015)}]{YoshinoYamauchi2015}
{Yoshino}, A., \& {Yamauchi}, C. 2015, \mnras, 446, 3749

\end{thebibliography}

\end{document}